\documentclass[reprint, aps, prl, twocolumn]{revtex4-1}
\usepackage{amsmath,amssymb,braket,fancybox}
\usepackage{amsfonts}
\usepackage{graphicx}
\usepackage{comment}
\usepackage{ascmac}
\usepackage{ascmac}
\usepackage{amsthm,color}

\theoremstyle{definition}

\newtheorem*{theorem*}{Theorem}

\newtheorem*{definition*}{Definition}

\newcommand{\av}[1]{\overline{#1}}
\newcommand{\mc}[1]{\mathcal{#1}}
\newcommand{\mr}[1]{\mathrm{#1}}

\newcommand{\mbf}[1]{\mathbf{#1}}

\newcommand{\lrs}[1]{\left( #1 \right)}

\newcommand{\lrl}[1]{\left[ #1 \right]}
\newcommand{\lrv}[1]{\left| #1 \right|}
\newcommand{\braketL}[1]{\left\langle #1 \right\rangle}

\newcommand{\fracd}[2]{\frac{\mathrm{d} #1 }{\mathrm{d} #2 }}

\newcommand{\fracpd}[2]{\frac{\partial #1 }{\partial #2 }}

\newcommand{\aln}[1]{
\begin{align}
#1
\end{align}
}

\newcommand{\ra}{\rightarrow}

\newcommand{\Tr}{\mr{Tr}}

\begin{document}
\title{
{Speed limits} to fluctuation dynamics
}
\date{\today}
\author{Ryusuke Hamazaki}
\affiliation{
Nonequilibrium Quantum Statistical Mechanics RIKEN Hakubi Research Team, RIKEN Cluster for Pioneering Research (CPR), RIKEN iTHEMS, Wako, Saitama 351-0198, Japan\\
email: ryusuke.hamazaki@riken.jp
}

\begin{abstract}
\textbf{
Fluctuation dynamics of an experimentally measured observable offer a primary signal for nonequilibrium systems, along with dynamics of the mean.
While universal speed limits for the mean have actively been studied recently,  
constraints for the speed of the fluctuation 
have been elusive.
Here, we develop a theory concerning rigorous limits to  the rate of fluctuation growth.
We find a principle that the speed of an observable’s fluctuation is upper bounded by the fluctuation of an appropriate observable describing velocity, which also indicates a tradeoff relation between the changes for the mean and fluctuation.
{We demonstrate the advantages of our inequalities for processes with non-negligible dispersion of observables, quantum work extraction, and the entanglement growth in free fermionic systems.}
Our results open an avenue toward a quantitative theory of fluctuation dynamics in various non-equilibrium systems encompassing quantum many-body systems and nonlinear population dynamics.
}

\end{abstract}

\maketitle
{\textbf{\large{Introduction}}}

Many physical systems involve randomness caused by stochastic noises due to external environment or intrinsic quantum uncertainty{, which have been central objects in nonequilibrium statistical mechanics~\cite{kubo1966fluctuation,seifert2012stochastic,RevModPhys.81.1665,PhysRevLett.114.158101,horowitz2020thermodynamic}.
Two primary quantities that characterize such fluctuating systems are the mean and the fluctuations of an experimentally measured observable.
When the system is out of equilibrium, the distribution of the fluctuating observable changes in time, which causes the time evolution for the mean and the fluctuation.
To understand nonequilibrium dynamics, rigorous bounds concerning the rate of the change for the mean value of the observable have intensively been studied recently~\cite{PhysRevE.97.062101,dechant2018current,PhysRevX.10.021056,nicholson2020time,PhysRevA.106.042436,PhysRevX.12.011038,PRXQuantum.3.020319,gong2022bounds}, though the first work on such ``speed limits~\cite{mandelstam1945energy,margolus1998maximum,PhysRevLett.110.050402,PhysRevLett.111.010402,PhysRevLett.110.050403,PhysRevX.6.021031,PhysRevLett.121.070601,PhysRevE.102.062132,PhysRevA.103.022210,PhysRevLett.130.010402,hasegawa2023unifying,bason2012high,ness2021observing,deffner2017quantum}" dates back more than a half-century ago by Mandelstam and Tamm~\cite{mandelstam1945energy}.
For example, the speed (i.e., the time derivative) of the mean is known to be bounded by a product of the standard deviation and the Fisher information~\cite{PhysRevX.10.021056,nicholson2020time,PhysRevX.12.011038}, which is related to the Cram\'er-Rao inequality in information theory.
Such speed limits for observables can provide more experimentally relevant and even tighter constraints than those for the entire states~\cite{PhysRevX.12.011038,PRXQuantum.3.020319}.
}

{
Although the speed of the mean value has been investigated so far, previous speed limits cannot make a clear prediction to  fluctuations  of an observable, e.g.,  the standard deviation, which are not simply given as a mean of a state-independent observable (see Ref.~\cite{PRXQuantum.3.020319} for an exception).
Such fluctuation growth plays an equally (or, sometimes even more) important role  compared with the mean in characterizing out-of-equilibrium dynamics.
For example, the spreading of particles should be characterized by the fluctuation dynamics of the particles' position as well as the mean displacement~\cite{schneider2012fermionic,PhysRevLett.110.205301}; in such a process, previous bounds for the mean can only provide rather loose bounds.
As another example, fluctuation dynamics of macroscopic observables can relate with dynamics of entanglement~\cite{PhysRevLett.127.090601,PhysRevB.102.195132} and coherence~\cite{PhysRevLett.95.090401} in certain quantum systems, which are not understood from the mean value alone.
It is thus a pivotal open question to elucidate the dynamical behavior of fluctuations of observables.
}

\begin{figure*}
\begin{center}
\includegraphics[width=\linewidth]{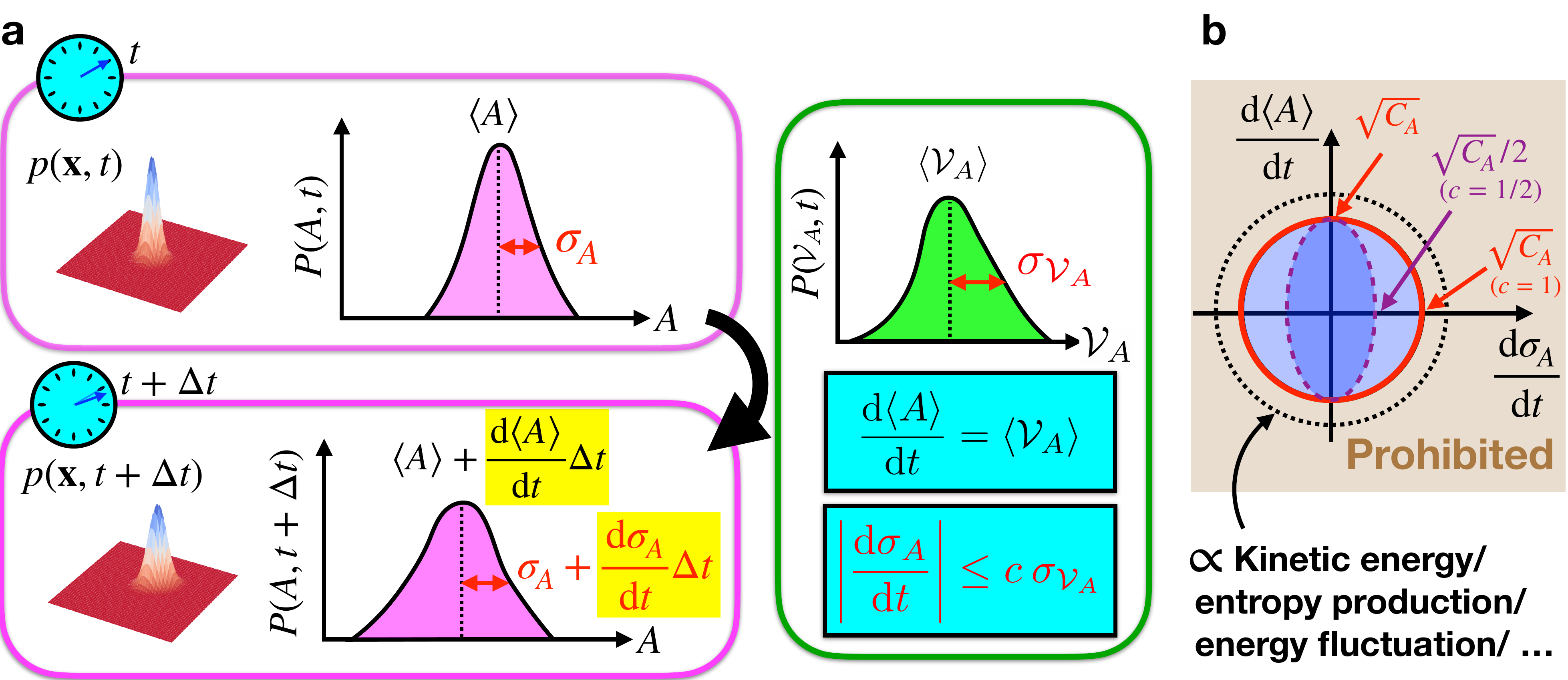}
\end{center}
\caption{
\textbf{Schematic illustrations of our limits to fluctuation dynamics.}
\textbf{a} The probability distribution at time $t$, $p(\mbf{x},t)$, defines the probability $P(A,t)$ for a random observable $A$.
After a short-time interval $\Delta t$, the mean $\braket{A}$ and the standard deviation $\sigma_A$ change by $\sim \fracd{\braket{A}}{t}\Delta t$ and $\sim{\fracd{\sigma_A}{t}}\Delta t$, respectively.
We define a suitable velocity observable $\mc{V}_A$, which satisfies $\fracd{\braket{A}}{t}=\braket{\mc{V}_A}$. Then, the speed of the fluctuation of $A$ is bounded by the fluctuation of the velocity as {$\lrv{\fracd{\sigma_A}{t}}\leq c\:\sigma_{\mc{V}_A}$, where $c=1$ or $c=1/2$.}
\textbf{b} The bound leads to the tradeoff relation \eqref{cost_main} between $\fracd{\braket{A}}{t}$ and $\fracd{\sigma_A}{t}$, i.e., they should be within the solid circle with magenta {for $c=1$. 
For $c=1/2$, we can tighten the bound,} where the allowed region becomes inside the dashed ellipse. As summarized in Table I,  {the sum of the squares for $\fracd{\braket{A}}{t}$ and $\frac{1}{c}\fracd{\sigma_A}{t}$} can be further bounded using physical quantities, e.g., the kinetic energy.
}
\label{sche}
\end{figure*}

Here, we develop a  theory for rigorous bounds on dynamics of fluctuations {of observables} (Fig.~\ref{sche}\textbf{a}).
We argue that {for a class of dynamics, the standard deviation of an observable $A$} has a speed always smaller than the standard deviation of the suitably chosen ``velocity observable" $\mathcal{V}_A$ of $A$.
This inequality also indicates a tradeoff relation between the speeds of the mean and the fluctuation.
Our inequalities are applicable to a wide range of nonequilibrium systems, ad shown in Table I.
Furthermore, we demonstrate that our bounds can be practically advantageous in concrete physical situations, such as  the transition accompanied by non-negligible dispersion of observables and the entanglement-entropy dynamics for non-interacting fermions. 
\newline

{\textbf{\large{Results}}}

\textbf{Speed limits to fluctuation dynamics.}
We argue that {for certain dynamics, the speed of the standard deviation of an observable $A$} is bounded by the standard deviation of the appropriately defined velocity observable $\mathcal{V}_A$, i.e.,
{
\aln{\label{fluc_main}
\lrv{\fracd{{\sigma}_A}{t}}\leq c\:\sigma_{\mathcal{V}_A}.
}
}
Here, $\sigma_A$ denotes the standard deviation of $A$,
and the velocity observable is chosen such that  $\fracd{\braket{{A}}}{t}=\braket{\mathcal{V}_A}$ is satisfied, where $\braket{\cdots}$ denotes the mean. 
{The constant $c$ is determined by the dynamics and the choice of $\mc{V}_A$, and we will obtain $c=1$ or $c=1/2$ in this manuscript.}
In addition, this inequality  indicates a tradeoff relation between the speeds of the mean and the standard deviation.
That is, two quantities cannot be simultaneously fast, which we concisely represent as (Fig.~\ref{sche}\textbf{b})
{
\aln{\label{cost_main}
\lrs{\fracd{\braket{{A}}}{t}}^2+\frac{1}{c^2}\lrs{\fracd{{\sigma}_A}{t}}^2\leq \braket{\mc{V}_A^2} =:C_A.
}
We call the equivalent inequalities~\eqref{fluc_main} and \eqref{cost_main}, as well as their underlying equality~\eqref{vel_eqm} below, as our main results.
Note that an inappropriate choice of $\mathcal{V}_A$, say $\mathcal{V}_A=\fracd{\braket{{A}}}{t}1$, breaks Eqs.~\eqref{fluc_main} and~\eqref{cost_main}.
Nonetheless, we will show that we can choose appropriate velocity observables where Eqs.~\eqref{fluc_main} and~\eqref{cost_main} hold for various physically relevant setups.}

{
When $C_A$ is evaluated easily (see, e.g., unitary quantum dynamics with~\eqref{unit_dyn}),  \eqref{cost_main} is already practically useful because we can upper bound the speeds of the mean and fluctuation.
Moreover, as another nontrivial result of ours, we discover that $C_A$ can be further upper bounded using other physical quantities for certain setups, such as the kinetic energy in classical and quantum hydrodynamics and the entropy production rate for thermodynamic processes.}
Table~I summarizes the broad applicability of our results.

\begin{table*}
{
 \caption{{\textbf{Summary of physical situations for \eqref{fluc_main}-\eqref{vel_eqm}.} We show the values of $c$, velocity observable, and the  quantity that upper bounds $\lrs{\fracd{\braket{{A}}}{t}}^2+\frac{1}{c^2}\lrs{\fracd{{\sigma}_A}{t}}^2$ in each non-equilibrium system. 
 For $c=1/2$, our results \eqref{fluc_main}-\eqref{vel_eqm} can be shown only for time-independent observables. The upper bound on $\lrs{\fracd{\braket{{A}}}{t}}^2+\frac{1}{c^2}\lrs{\fracd{{\sigma}_A}{t}}^2$ in the rightmost column is also for time-independent observables.}}}
 \centering
  \begin{tabular}{|c|c|c|c|}
   \hline
Non-equilibrium system& $c$ & Velocity observable & Quantity that bounds $\lrs{\fracd{\braket{{A}}}{t}}^2+\frac{1}{c^2}\lrs{\fracd{{\sigma}_A}{t}}^2$ \\
   \hline
   Classical and quantum hydrodynamics & 1& $\nabla A(\mathbf{x})\cdot \mbf{u}(\mathbf{x})+\dot{A}(\mbf{x})$ & Kinetic energy $E_\mr{kin}$ \\
   \hline
Fokker-Planck dynamics & 1& $\nabla A(\mathbf{x})\cdot \mbf{v}(\mathbf{x})+\dot{A}(\mbf{x})$ &Entropy production rate $\dot{\Sigma}$ \\
   \hline
Macroscopic transport in discrete quantum systems& 1& $\frac{1}{2}(\nabla A)_{ij} V_{ij}+\dot{A}_i $& Modified transition strength ${\mathcal{S}_H^2-E_\mr{trans}^2}$ \\
\hline
Evolutionary dynamics without mutation& 1/2& $(\delta A)_i\:(\delta s)_i $& Variance of the growth rate $\sigma_s^2$ \\
   \hline
General (non)linear dynamics& 1/2& $-(\delta A)_i \dot{I} $& Classical Fisher information $\mathcal{F}_\mr{C}$ \\
   \hline
   Unitary quantum dynamics& 1& $i[\hat{H},\hat{A}]/\hbar+\dot{\hat{A}}$ & $-$ \\
   \hline
      Unitary quantum dynamics& 1/2& $-$ & (Energy variance ${\Delta E^2}$) \\
   \hline

      Dissipative quantum dynamics&\:1/2\: & $-$ & (Quantum Fisher information $\mc{F}_\mr{Q}$) \\
   \hline
  \end{tabular} \label{AUSclass}
\end{table*}

{
\textbf{General proof and the required condition.}
We first provide a general proof of our main results~\eqref{fluc_main}
and \eqref{cost_main}. We especially identify nontrivial sufficient conditions for our inequalities to hold, with discussing how to choose the velocity observable. 
For this purpose, we first show the following important equality for the variance of $A$,
\aln{\label{vel_eqm}
\fracd{\braket{\delta A^2}}{t}=2c\braket{\delta A,\delta \mc{V}_A},
}
where $\delta X=X-\braket{X}$ is the fluctuation observable and $\braket{X,Y}=\frac{1}{2}\braket{\{X,Y\}}$ is the correlation between $X$ and $Y$.
Here, we have introduced the anti-commutator $\{X,Y\}=XY+YX$ to include the quantum case with non-commutativity between $X$ and $Y$ ($\{X,Y\}=2XY$ for classical cases).
Note that the right-hand side in Eq.~\eqref{vel_eqm} is given as the covariance between $A$ and $\mc{V}_A$.
If we obtain \eqref{vel_eqm}, \eqref{fluc_main} can be shown by the Cauchy-Schwarz inequality $|\braket{\delta A,\delta \mc{V}_A}|\leq \sigma_A\sigma_{\mc{V}_A}$ and $\fracd{\braket{\delta A^2}}{t}=2\sigma_A\fracd{\sigma_A}{t}$.
}

{
We now discuss conditions for Eq.~\eqref{vel_eqm} to hold and ansatz to choose $\mc{V}_A$. For this purpose, we introduce a notation where the expectation value of an observable $A$ is given by some inner product between $A$ and some probability (density) $\rho$ as $\braket{A}=(A|\rho)$. 
For example, for continuous classical systems we have a probability density $\rho=\{\rho(\mbf{x},t)\}$ and $(A|\rho)=\int d\mbf{x}A(\mbf{x},t)\rho(\mbf{x},t)$; for classical discrete systems we have a probability distribution $\rho=\{p_i(t)\}$ and 
$(A|\rho)=\sum_iA_i(t)p_i(t)$; for quantum systems, we have a density matrix 
$\rho=\hat{\rho}(t)$ and 
$(A|\rho)=\Tr[\hat{A}(t)\hat{\rho}(t)]$.
Now, we assume that the time evolution of the probability is given by
$
\fracd{\rho}{t} =\mc{L}[{\rho}],
$
where the map $\mc{L}$ is not unique and can depend on $\rho$.
We can also define the dual map $\mc{L}^\dag$ of $\mc{L}$ satisfying the relation
$
(A|\dot{\rho})=(A|\mc{L}[\rho])=(\mc{L}^\dag[A]|\rho),
$
which is required to hold for an arbitrary  observable $A$.
}

{
The above definition for $\mc{L}^\dag$ leads to the natural way to find the velocity observable. Indeed, if we define
\aln{\label{velans}
\mc{V}_A=\dot{A}+\mc{L}^\dag[A],
} 
$\mc{V}_A$ satisfies the requirement for the velocity observable because
\aln{\label{mean_m}
\fracd{\braket{A}}{t}=(\dot{A}|\rho)+(A|\dot{\rho})=(\mc{V}_A|\rho)=\braket{\mc{V}_A},
}
where $\dot{X}=\fracd{X}{t}$ (see \textbf{Methods} for the possible freedom of choosing velocity observables). As discussed later, Eq.~\eqref{mean_m} with \eqref{velans} has often been used to derive speed limits for the mean of $A$~\cite{nicholson2020time,PhysRevX.10.021056,PhysRevX.12.011038,PRXQuantum.3.020319}. Indeed, by finding the further upper bound for $|\braket{\mc{L}^\dag[A]}|$, one obtains the speed limit for $\lrv{\fracd{\braket{A}}{t}-\dot{A}}$ in terms of, e.g., Fisher information~\cite{nicholson2020time,PhysRevX.10.021056,PhysRevX.12.011038}.
However, these previous results cannot lead to Eq.~\eqref{vel_eqm} and our main bounds \eqref{fluc_main} and \eqref{cost_main}.
}

{
To derive Eq.~\eqref{vel_eqm} under the choice in \eqref{velans}, we need an additional nontrivial condition. Indeed, a straightforward calculation leads to
$\fracd{\braket{\delta A^2}}{t}=(\{\delta A,\dot{A}\}+\mc{L}^\dag[A^2]-2\braket{A}\mc{L}^\dag[A]|\rho)$.
We thus need
\aln{\label{crureq_m}
\braket{\{\delta A,\dot{A}\}+\mc{L}^\dag[A^2]-2\braket{A}\mc{L}^\dag[A]}=c\braket{\{\delta A,\mc{V}_A\}}
}
to obtain \eqref{vel_eqm} (note that $\braket{\{\delta A,\mc{V}_A\}}=\braket{\{\delta A,\delta\mc{V}_A\}}$). We stress that this condition is unique to  our analysis for fluctuation dynamics, which goes beyond previous speed limits for the mean~\cite{nicholson2020time,PhysRevX.10.021056,PhysRevX.12.011038,PRXQuantum.3.020319}.
}

{
We find two different situations \eqref{fsit} and \eqref{sfit}, each of which ensures \eqref{crureq_m}.
The first situation corresponds to
\aln{\label{fsit}
\mc{L}^\dag[A^2]=\{A,\mc{L}^\dag[A]\},
}
which leads to Eq.~\eqref{crureq_m} with $c=1$. 
Importantly, if $\mc{L}^\dag$ represents a derivative in an abstract sense, i.e.,
 $\mc{L}^\dag$ is linear and satisfies the Leibniz rule $\mc{L}^\dag[AB]=A\mc{L}^\dag[B]+\mc{L}^\dag[A]B$, Eq.~\eqref{crureq_m} holds.
}

\textbf{Limits from the local conservation law.}
{There are several class of dynamics satisfying \eqref{fsit}. As a first class, } let us consider continuous dynamics under local conservation law of some normalized distribution {$\rho=\{\rho(\mbf{x},t)\}$} with $\int d\mbf{x}\rho(\mbf{x},t)=1$.
{It is then natural to define the mean and the standard deviation of a space-dependent observable $A$ coupled to $\rho$ as $(A|\rho)=\braket{A(t)}= \int d\mbf{x} A(\mbf{x})\rho(\mbf{x},t)$
and $\sigma_A(t)= \sqrt{\int d\mbf{x} (A(\mbf{x})-\braket{A(t)})^2\rho(\mbf{x},t)}$, respectively.}

We assume that the continuity equation {$\partial_t\rho(\mbf{x},t)=\mc{L}[\rho]=-\nabla\cdot (\rho(\mbf{x},t)\mbf{V}(\mbf{x},t))$}
holds by introducing the normalized current $\mbf{V}(\mbf{x},t)$, and that  $\rho\mbf{V}$ vanishes in the limit $|\mbf{x}|\ra\infty$.
{Using the integration by parts, we can find $\mc{L}^\dag[A]=\nabla A\cdot \mathbf{V}$ and thus
\aln{\label{velobscon}
\mc{V}_A(\mbf{x},t)=\nabla A(\mbf{x},t)\cdot \mbf{V}(\mbf{x},t) +\dot{A}(\mbf{x},t).
}
by using Eq.~\eqref{velans}.
}

{Now, since $\nabla A^2=2A\nabla A$, Eq.~\eqref{fsit}  holds for $\mc{L}^\dag$, and thus Eq.~\eqref{crureq_m} and Eq.~\eqref{vel_eqm} do.
Therefore, our main bounds on the fluctuation dynamics, \eqref{fluc_main} and \eqref{cost_main}, are obtained for the velocity observable \eqref{velobscon} with $c=1$.}

{
Now, if we consider time-independent observables, we can further find the upper bound of $C_A=\braket{(\nabla A\cdot \mbf{V})^2}$ by virtue of, e.g., the H\"older's inequality:
\aln{\label{cost_ev}
C_A\leq \|\nabla A\|_\infty^2 \cdot\int d\mbf{x}\rho(\mbf{x},t)|\mbf{V}(\mbf{x},t)|^2.
}
Here, $\|\nabla A\|_\infty=\max_\mbf{x}|\nabla A(\mbf{x})|$, which is calculated easily for a given $A$.
We stress that this result cannot be obtained from the previous speed limit for the mean despite the appearance of a similar factor. 
Indeed, the previous approach~\cite{PhysRevE.97.062101,PRXQuantum.3.020319} uses the Cauchy-Schwarz inequality to obtain $\lrv{\fracd{\braket{A}}{t}}^2=|\braket{\mc{L}^\dag[A]}|^2=|\braket{\nabla A\cdot \mbf{V}}|^2\leq \braket{(\nabla A)^2}\int d\mbf{x}\rho(\mbf{x},t)|\mbf{V}(\mbf{x},t)|^2$, where we do not need the condition~\eqref{fsit} for the proof.
}

Let us provide some examples of our general argument above:

\textit{{Example 1}: Irreversible thermodynamics.}
Let us consider thermodynamic systems of Brownian particles described by the overdamped Fokker-Planck equation~\cite{seifert2012stochastic}.
In this case, the probability density {$\rho(\mbf{x},t)$} satisfies the continuity equation as $\partial_t\rho=-\nabla\cdot(\rho\mathbf{v})=-\nabla\cdot(\rho\mu(\mathbf{F}-T\nabla \ln\rho))$, where $\mbf{v}$ is the local velocity for the probability current, $\mu$ is the mobility, $T$ is the temperature, and $\mbf{F}$ is the force. 
Then, our results Eqs.~\eqref{fluc_main}-\eqref{vel_eqm} hold for any function $A(\mbf{x},t)$ with $\mc{V}_A=\nabla A\cdot\mbf{v}+\dot{A}$.
For time-independent observables, $C_A$  is bounded by the total entropy production rate~\cite{seifert2012stochastic}  $\dot{\Sigma}=\frac{1}{\mu T}\int d\mbf{x}\rho|\mbf{v}|^2$ as $C_A\leq \mu T\|\nabla A\|_\infty^2 \dot{\Sigma}$.
Thus, to change the mean and the fluctuation quickly, the irreversibility of the system should be significant; note that the previous classical speed limits only considered the speed of the mean~\cite{PhysRevE.97.062101,dechant2018current,PhysRevLett.121.070601}.

\begin{figure}
\begin{center}
\includegraphics[width=\linewidth]{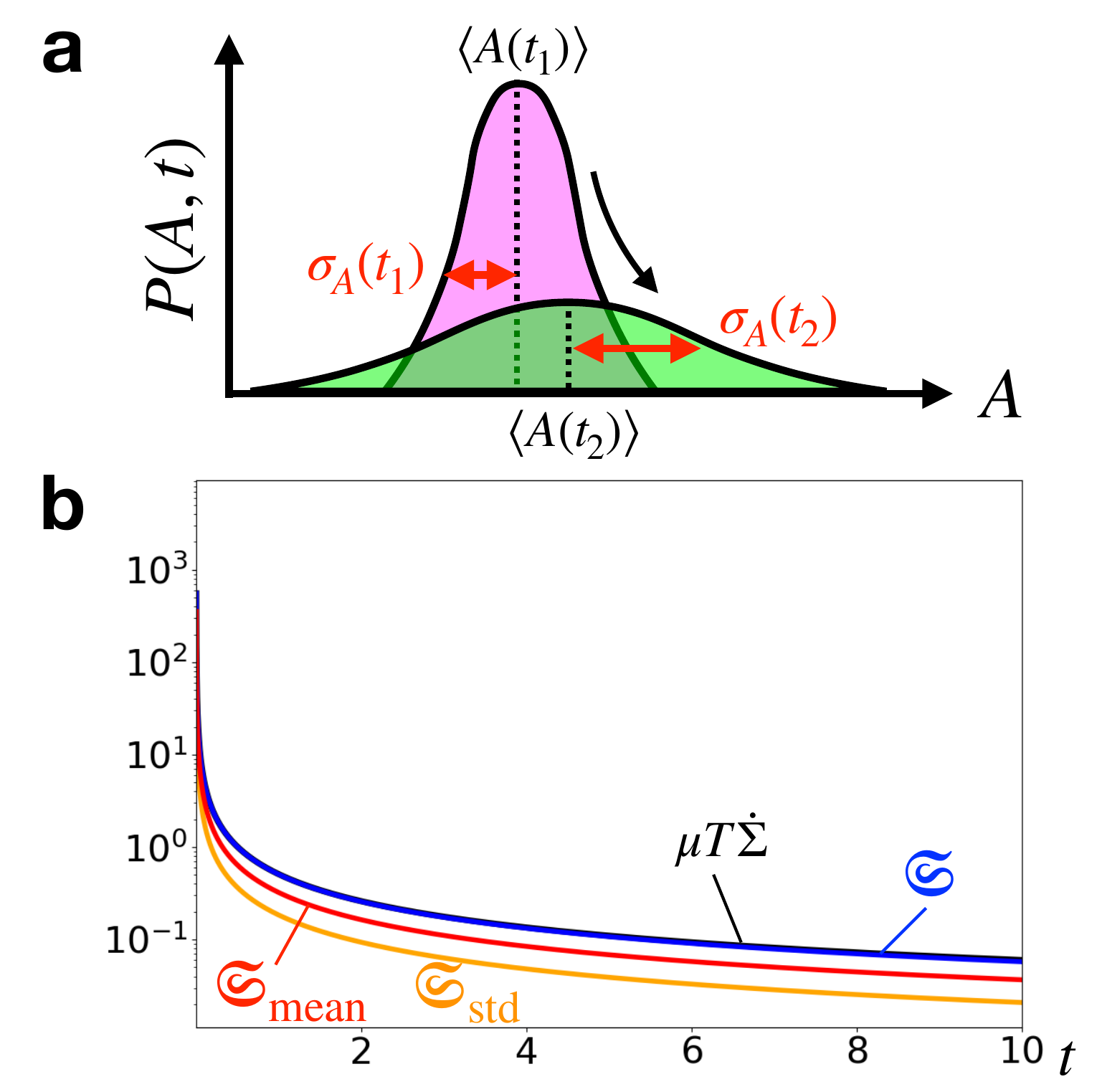}
\end{center}
\caption{
{
\textbf{Transition with non-negligible dispersion.}
\textbf{a} Schematic illustration for the transition of the observable $A$. When the change of the standard deviation $\sigma_A$ for time $[t_1,t_2]$ is comparable to or larger than that for the mean $\braket{A}$, our results become significant.
\textbf{b} Time evolution of the (scaled) entropy production rate $\mu T\dot{\Sigma}$ (black),
$\mathfrak{S}$ in \eqref{now2} (blue), $\mathfrak{S}_\mr{mean}=\lrs{\fracd{\braket{A}}{t}}^2$ (red), and $\mathfrak{S}_\mr{std}=\lrs{\fracd{\sigma_A}{t}}^2$ (orange) with $A=|x|$
for the one-dimensional Fokker-Planck equation $\partial_t\rho=-\partial_x(\rho\mu(F-T\partial_x\ln \rho))$.
Our result  in \eqref{now2}  provides a considerably tight bound concerning $\mathfrak{S}$ and $\mu T\dot{\Sigma}$, where black and blue lines almost overlap, while previous bounds  in $\mathfrak{S}_\mr{mean}\leq \mu T\dot{\Sigma}$ and  $\mathfrak{S}_\mr{std}\leq \mu T\dot{\Sigma}$ do not. 
We use $\mu=T=1$, $F=0.1$ (constants), and the initial distribution $\rho(x,t=0)=\delta(x)$.}
}
\label{application_ent_m}
\end{figure}

{
Our bounds can become advantageous over previous ones if we consider a transition with non-negligible dispersion of observables as shown in Fig.~\ref{application_ent_m}\textbf{a}.
In this case, we typically have $\lrv{\fracd{\braket{A}}{t}}\sim \lrv{\fracd{\sigma_{A}}{t}}$ or $\lrv{\fracd{\braket{A}}{t}}\ll \lrv{\fracd{\sigma_{A}}{t}}$ at an instantaneous time.
Then, the previous works on the speed limit for the mean of observables can be problematic in understanding such a transition, i.e., they can provide rather loose bounds.
In contrast, our inequality~\eqref{cost_main} succeeds in resolving this problem by considering the degree of dispersion of the distribution.}

{
As a simple demonstration, let us consider 
 an example of a classical particle obeying the Fokker-Planck equation and take the distance of the particle from the origin, $A(\mathbf{x})=|\mathbf{x}|$, as an observable.
We consider a localized initial state at the origin, for which $\braket{A(0)}$ and $\sigma_A(0)$ are very small.
Due to the time evolution,  $\braket{A}$ and $\sigma_A$ change, but since we assume that the diffusion is non-negligible, the rate $\lrv{\fracd{\sigma_A}{t}}$ is comparable to 
$\lrv{\fracd{\braket{A}}{t}}$.
Now, the previously known inequality~\cite{ito2023geometric} leads to
$
\mathfrak{S}_\mr{mean}:=\lrs{\fracd{\braket{A}}{t}}^2\leq \mu T\dot{\Sigma}.
$
Moreover, using the method in our previous paper~\cite{PRXQuantum.3.020319}, we also find
$\mathfrak{S}_\mr{std}:=\lrs{\fracd{\sigma_{A}}{t}}^2\leq \mu T\dot{\Sigma},
$
where we have used the fact $\|\nabla A\|_\infty=1$. 
However, these inequalities are far from tight, as numerically demonstrated in the red and orange curves for a one-dimensional system in Fig.~\ref{application_ent_m}\textbf{b}.
}

{
In stark contrast, the inequality obtained from our result,
\aln{\label{now2}
\mathfrak{S}:=\lrs{\fracd{\braket{A}}{t}}^2+\lrs{\fracd{\sigma_{A}}{t}}^2\leq \mu T\dot{\Sigma},
}
can lead to a considerably tight result, as shown in the blue curve of Fig.~\ref{application_ent_m}\textbf{b}.
{While weconsider a simple observable $A=|\mbf{x}|$ to demonstrate our general result  here,
 this inequality still applies to more general time-independent observables by changing the right-hand side into $\|\nabla A\|_\infty^2 \mu T\dot{\Sigma}$.}}
Therefore, our inequality is useful to evaluate $\lrv{\fracd{\sigma_A}{t}}$ and 
$\lrv{\fracd{\braket{A}}{t}}$ simultaneously when we know the entropy production rate for a general observable.
In turn, when we can measure $\lrv{\fracd{\sigma_A}{t}}$ and 
$\lrv{\fracd{\braket{A}}{t}}$ but not $\dot{\Sigma}$, our inequality is useful for tightly evaluating the entropy production rate from the dynamics of a given observable $A$. 
{Note that this result can be also derived from the infinitesimal-time limit of the inequality concerning the Wasserstein distance, where we use some known results in, e.g., Ref.~\cite{dechant2019thermodynamic}
 (also see \textbf{Methods.}). }

\textit{{Example 2}: Classical and quantum hydrodynamics.}
Consider the classical hydrodynamics described by, e.g., the nonlinear Navier-Stokes equation.
For the density $\rho(\mbf{x},t)$ of the fluid, we generally have a continuity equation $\partial_t\rho=-\nabla\cdot(\rho\mbf{u})$, where $\mbf{u}$ is the current of the fluid.
Similarly, we can consider quantum hydrodynamics~\cite{tsubota2013quantum} that is described by the nonlinear Schr\"{o}dinger equation. While the quantum fluid is described by the complex amplitude $\Psi(\mbf{x},t)$, its density $\rho(\mbf{x},t)=|\Psi(\mbf{x},t)|^2$ obeys the continuity equation $\partial_t\rho=-\nabla\cdot(\rho\mbf{u})$.
Then, in both cases, our results Eqs.~\eqref{fluc_main}-\eqref{vel_eqm} hold for any function given by $A(\mbf{x},t)$ with $\mc{V}_A=\nabla A\cdot\mbf{u}+\dot{A}$.
For time-independent observables, $C_A$ is bounded by the kinetic energy $E_\mr{kin}=\frac{M}{2}\int d\mbf{x}\rho|\mbf{u}|^2$ as $C_A\leq \frac{2\|\nabla A\|_\infty^2 E_\mr{kin}}{M}$, where $M$ is the total mass. 

{Our bound has a physical meaning that the speed of mean and fluctuation becomes small simultaneously if the kinetic energy $E_\mr{kin}$ is small, irrespective of potential and interaction energies. In addition, our bound again becomes much tighter than that in \cite{PRXQuantum.3.020319} when we prepare a situation satisfying $\lrv{\fracd{\braket{A}}{t}}\sim \lrv{\fracd{\sigma_{A}}{t}}$, in a manner similar to \textit{Example 1}.}

{
Before ending this section, we make several remarks, which are discussed in \textbf
{Methods} in detail.
First, while the above discussions are for standard deviations,
we further discover that  inequalities similar to that in \eqref{fluc_main} are obtained for 
even higher-order absolute central  moments, 
$\mu_{A}^{(n)}:=\braket{|A-\braket{A}|^n}$. Using the local conservation law of probability, we prove
$
\lrv{\fracd{(\mu_{A}^{(n)})^{\frac{1}{n}}}{t}}\leq (\mu_{\mc{V}_A}^{(n)})^{\frac{1}{n}},
$
which reduces to \eqref{fluc_main} for $n=2$.
Second, we note that 
the bound on $C_A$, \eqref{cost_ev}, is related to the Wasserstein geometry in the optimal transport theory~\cite{villani2009optimal}.
Indeed, by minimizing the right-hand side of \eqref{cost_ev} in terms of $\mbf{V}$ appearing in the continuity equation, we find~\cite{benamou2000computational,PhysRevResearch.3.043093} that $C_A$ is further upper bounded by a factor involving the order-2 Wasserstein distance.
However, our general results in~\eqref{fluc_main} and \eqref{cost_main}
are tighter and not directly understood from  the Wasserstein geometry.
Third, though we have considered continuous systems, a similar structure appears for discrete systems that satisfy the discrete version of the local conservation law of probability. While the details are given in \textbf{Methods}, we here mention that this leads to a tighter bound for the transport in quantum many-body systems than those obtained before~\cite{PRXQuantum.3.020319,PhysRevLett.130.010402} (see \textit{Example 6} in \textbf{Methods} and Supplementary Note 1).
}

\textbf{Limits in unitary quantum dynamics.}
{Our main results in \eqref{fluc_main}, \eqref{cost_main}, and \eqref{vel_eqm} with $c=1$ also hold for a completely different situation, i.e., unitary quantum dynamics, described by the von Neumann equation of the density matrix as $\fracd{\hat{\rho}}{t}=-\frac{i}{\hbar}[\hat{H},\hat{\rho}]$.
Indeed, we can take $\mc{L}^\dag[\hat{A}]=\frac{i}{\hbar}[\hat{H},\hat{A}]$, which satisfies \eqref{fsit} and thus \eqref{crureq_m}.
Therefore, our main results apply under the velocity observable
\aln{
\hat{\mc{V}}_A=\frac{i}{\hbar}[\hat{H},\hat{A}]+\dot{\hat{A}}.
}
For example, for time-independent $\hat{A}$,~\eqref{cost_main} reads
\aln{\label{unit_dyn}
\lrs{\fracd{\braket{\hat{A}}}{t}}^2+\lrs{\fracd{\sigma_A}{t}}^2\leq C_A=-\frac{\braket{[\hat{H},\hat{A}]^2}}{\hbar^2},
}
where $C_A$ is often easy to evaluate (see the time-dependent case later). This inequality is meaningful for the semi-classical limit $\hbar\ra 0$, in stark contrast with the Mandelstam-Tamm bound~\cite{mandelstam1945energy} (see \textbf{Methods}).
Note that  $\mc{\hat{V}}_A$ is the Schrodinger picture of the time derivative of the observable in the Heisenberg representation.
Indeed, its Heisenberg picture satisfies $\mc{\hat{V}}_A^\mathrm{H}(t)=\fracd{\hat{A}^\mathrm{H}(t)}{t}$ with $\hat{X}^\mathrm{H}(t)=U^\dag(t)\hat{X}(t)U(t)$ and $U(t)=\mathbb{T}e^{-i\int_0^t\hat{H}(s)ds}$ ($\mathbb{T}$ denotes the time ordering).
}

{\textit{Example 3: Dynamics of spin systems.}}
We first consider $\hat{H}= g\hat{s}^x$ and $\hat{A}=\hat{s}^z$, where $\hat{s}^{s,y,z}$ is the spin-1/2 spin operator. 
Let us take an initial state given by $\ket{\psi_0}=\cos(\theta/2)\ket{\uparrow}+i\sin(\theta/2)\ket{\downarrow}$ with arbitrary $\theta\in\mathbb{R}$, where $\ket{\uparrow}\:(\ket{\downarrow})$ is the eigenstate of $\hat{s}^z$ with eigenvalue $+1\:(-1)$.
We then find that  $\lrs{\fracd{\braket{{\hat{A}}}}{t}}^2+\lrs{\fracd{\sigma_A}{t}}^2=-\braket{[\hat{H},\hat{A}]^2}/\hbar^2$ holds for arbitrary time, which means that inequalities \eqref{fluc_main} and \eqref{cost_main} become equalities. 

Our results are useful even for quantum many-body systems.
For example, let us consider a spin-$S$ system whose Hamiltonian reads
$\hat{H}=\sum_{ij}J_{ij}(\hat{S}_i^x\hat{S}_j^x+\hat{S}_i^y\hat{S}_j^y)+\Delta_{ij}\hat{S}_i^z\hat{S}_j^z+\sum_ih_i\hat{S}_i^z+g\hat{S}_i^x$, where $J_{ij},h_i$ and $g$ are arbitrary ($\hat{S}_i^{x,y,z}$ is the spin-$S$ spin operator at site $i$).
Then, if we take $\hat{A}=\hat{M}_z=\sum_i\hat{S}_i^z$, we have 
$\lrv{\fracd{{\sigma}_{M_z}}{t}}\leq |g|\sigma_{M_y}$, where $\hat{M}_y=\sum_i\hat{S}_i^y$.
Therefore, fluctuation dynamics of magnetization in $z$-direction is simply bounded by the fluctuation of that in $y$-direction (or vice versa). 
We numerically demonstrate our limits to fluctuation dynamics for quantum many-body systems in Supplementary Note 2.
{There, we also show that our bound leads to the tighter estimation of the speed of the mean than the Mandelstam-Tamm bound.}

{\textit{Example 4: Amount of work done.}
Importantly, our bounds are also applicable to time-dependent observables. This fact leads to a  speed limit for the amount of work done in time-dependent unitary dynamics. 
Specifically, let us assume the initial state $\hat{\rho}(0)$ that is diagonal in the energy eigenbasis of $\hat{H}(0)$.
In this case, the instantaneous averaged work done on the system at time $t$ is represented by $\fracd{{W}(t)}{t}=\fracd{\braket{\hat{H}(t)}}{t}=\mr{Tr}\lrl{\hat{\rho}(t)\fracd{\hat{H}(t)}{t}}$, and the total work is given by $W(t_\mathrm{fin})=\Tr[\hat{\rho}(t_\mathrm{fin})\hat{H}(t_\mathrm{fin})]-\Tr[\hat{\rho}(0)\hat{H}(0)]$~\cite{PhysRevE.94.010103}.
We also find that $\hat{\mc{V}}_A=\fracd{\hat{H}(t)}{t}$. Then, our inequality leads to 
\aln{
\lrs{\fracd{{W}}{t}}^2+\lrs{\fracd{\sigma_{H(t)}}{t}}^2\leq C_{H(t)}=\braketL{\lrs{\fracd{\hat{H}(t)}{t}}^2}.
}
This inequality indicates that, given $C_{H(t)}$, both the instantaneous work and the fluctuation change are upper-bounded.
Note that we often consider $\hat{H}(t)=\hat{H}_1+\lambda(t)\hat{H}_2$, where $\hat{H}_2$ takes much simpler form than $\hat{H}_1$ (e.g., $\hat{H}_1$ has a complicated interaction but $\hat{H}_2$ is non-interacting).
In such cases, $C_{H(t)}$ is easier to access experimentally than the instantaneous work and energy fluctuation.}
{Moreover, since $C_{H(t)}\leq \lrv{\lrv{\fracd{\hat{H}(t)}{t}}}_\infty^2$, the above inequality states that the instantaneous work and the fluctuation change cannot be large when the control cost of changing the Hamiltonian $\lrv{\lrv{\fracd{\hat{H}(t)}{t}}}_\infty$ is given.}

{Furthermore, integrating over the entire interval and using the triangle inequality, we  have
\aln{\label{integ}
{W({t_\mr{fin}})^2+(\sigma_{H(t_\mr{fin})}-\sigma_{H(0)})^2}\leq t_\mr{fin}^2{\av{C_{H(t)}}},
}
where $\av{X}=\frac{1}{t_\mr{fin}}\int_0^{t_\mr{fin}}Xdt$.
There are some applications of these inequalities. For example, if the work and its fluctuation are difficult to measure because of complicated form of $\hat{H}$, we can use \eqref{integ} to evaluate it by $\av{C_H(t)}$, which is often easy to measure (see the previous paragraph).
As another implication, let us assume
that we want to perform a state transition from initial to final states with the given amount of work $W_\mr{fin}$ and the energy-fluctuation difference $\Delta \sigma_\mr{fin}$.
Then, \eqref{integ} immediately indicates the lower bound for the transition time in terms of $\av{C_H(t)}$, which quantifies the average magnitude of the Hamiltonian change,  as 
$
t_\mr{fin}\geq \sqrt{\frac{W_\mr{fin}^2+\Delta \sigma_\mr{fin}^2}{\av{C_{H(t)}}}}.
$
We note that even a weaker version of this inequality, $
t_\mr{fin}\geq \frac{|W_\mr{fin}|}{\sqrt{\av{C_{H(t)}}}}, 
$
has not been obtained before, while this is a fundamental speed limit concerning the work.
We also note that several
recent works discussed similar bounds on the speed of the erergy fluctuations, which set the quantum acceleration limits for the projective Hilbert space~\cite{pati2023quantum,cafaro2024uncertainty}.
}

\begin{figure}
\begin{center}
\includegraphics[width=\linewidth]{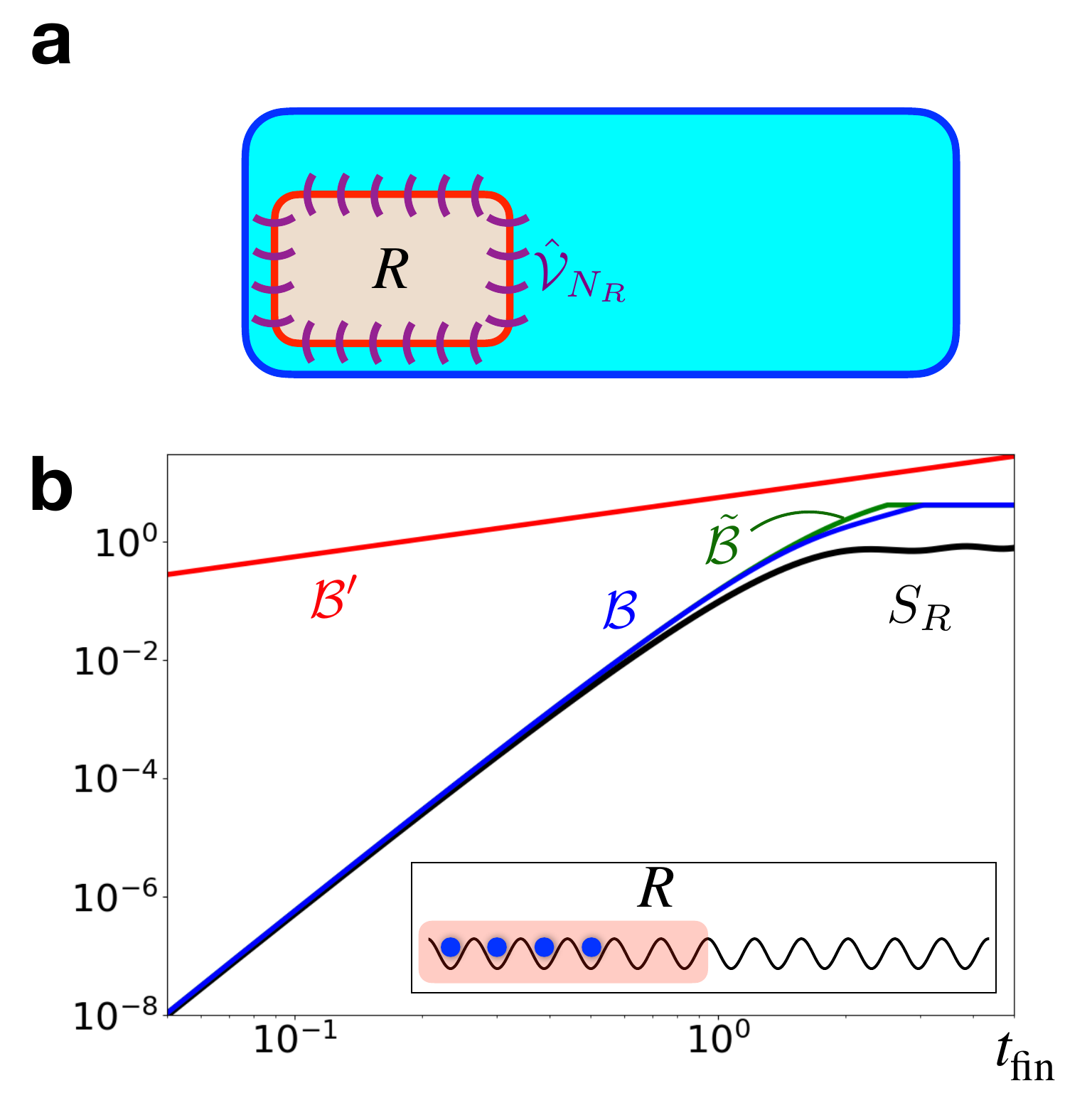}
\end{center}
\caption{
\textbf{Entanglement dynamics for non-interacting fermions.}
{
\textbf{a} Decomposition of a system into a subsystem $R$ and the rest, between which we consider the growth of the entanglement entropy.
For non-interacting fermions, entanglement growth is evaluated by measuring the fluctuation of the velocity observable $\mc{\hat{V}}_{N_R}=i[\hat{H},\hat{N}_R]$, where $\hat{N}_R$ denotes the number of particles in $R$. Importantly, $\mc{\hat{V}}_{N_R}$ only acts on the boundary of $R$.
\textbf{b} {The growth of entanglement entropy $S_R$ (black), our bounds $\mc{B}$ in \eqref{entbound} (blue) and $\mathcal{\tilde{B}}$ (green), and the previous bound $\mathcal{B}':=8(\ln 2)J_{k,k+1}t$ (red).
Our bounds provide good results, especially for the short-time regime, and become better than the previous bound.} We use a one-dimensional system with 12 sites and 4 fermions, and $R=\{1,2,3,4,5,6\}$.
The fermions are initially put on the lattice sites on the left.
We take $J_{i,i+1}=1$ and $V_i=0$ in Eq.~\eqref{fermodel}.}
}
\label{application_quant_m}
\end{figure}

{\textbf{Entanglement growth for non-interacting fermions.}
Our result can provide a useful rigorous bound for the growth of quantum entanglement~\cite{jurcevic2014quasiparticle,PhysRevX.7.031016,brydges2019probing,PhysRevLett.127.090601} for general non-interacting fermions on a lattice.
Here, let us especially consider the following Hamiltonian
$\hat{H}=\sum_{i,j}J_{ij}(\hat{c}_i^\dag \hat{c}_j +\mathrm{h.c.})+\sum_i V_i\hat{n}_i$,
where $\hat{c}_i$ and $\hat{n}_i=\hat{c}_i^\dag\hat{c}_i$ are the fermion annihilation and  number operator for site $i$, respectively, and $J_{ij}=J_{ji}$.
We consider the von Neumann entanglement entropy $S_R(t)=-\Tr[\hat{\rho}_R(t)\ln\hat{\rho}_R(t)]$ at a region $R$ composed of some lattice sites (Fig.~\ref{application_quant_m}\textbf{a}), where $\hat{\rho}_R$ is the reduced density matrix for $R$.
}

{
While it is not easy to experimentally measure the von Neumann entanglement entropy, we show that  it can be bounded by the number fluctuation at the region $R$ by developing the approach in Ref.~\cite{PhysRevLett.127.090601}.
Namely, if we define the number operator at region $R$, i.e., $\hat{N}_R=\sum_{i\in R}\hat{n}_i$, we show that 
\aln{\label{entineq}
S_R\leq \sigma_{N_R}^2\ln y +\frac{4|R|}{ey}
}
holds for the fermionic Gaussian states for arbitrary positive variable $y$, where $|R|$ denotes the number of sites in $R$ (see Supplementary Note 3 for the proof).
However, the right-hand side still requires the evaluation of the particle-number fluctuation for the entire bulk region in $R$.
}

{
Our limit for the fluctuation dynamics
\aln{\label{flucint}
(\sigma_{N_R}(t_\mathrm{fin})-\sigma_{N_R}(0))^2\leq \lrs{\int_0^{t_\mathrm{fin}}dt \sigma_{\mc{V}_{N_R}}}^2=:\mc{S},
}
which is obtained by integrating \eqref{fluc_main} over $t\in [0,t_\mr{fin}]$,
solves this problem since we can evaluate $\mc{S}$ and thus the entanglement entropy only from the quantities at the \textit{boundary} of $R$.
Specifically, let us assume a Gaussian state with no initial entanglement, for which $S_R(0)=\sigma_{N_R}(0)=0$.
Then, using \eqref{entineq} and a proper optimization of $y$, we discover that the entanglement entropy at time $t_\mathrm{fin}$ is upper bounded as (see \textbf{Methods})
\aln{\label{entbound}
S_{R}(t_\mathrm{fin})\leq \mc{B}:=
\left\{
\begin{array}{ll}
\mc{S}\:{\ln\frac{4|R|}{\mc{S}}} & \quad (\mc{S}\leq \frac{|R|}{4}),\\
{|R|\ln 2} &\quad (\mc{S}\geq \frac{|R|}{4}).
\end{array}
\right.
}
}

{
Strikingly, $\mc{V}_{N_R}=i[\hat{H},\hat{N}_R]$ only acts on the boundary of $R$, and thus $\sigma_{\mc{V}_{N_R}}$, $\mc{S}$, and $\mc{B}$ do not require the evaluation in the entire bulk region of $R$.
For a concrete example, consider a one-dimensional Hamiltonian with the next-neighbor hopping, i.e.,
\aln{\label{fermodel}
\hat{H}=\sum_{i=1}^LJ_{i,i+1}(\hat{c}_i^\dag \hat{c}_{i+1} +\mathrm{h.c.})+\sum_i V_i\hat{n}_i
}
and $R={1,2,\cdots,k}$.
Then, we can straightforwardly evaluate the bounds appearing in the above inequalities. 
For example, $\sigma_{\mc{V}_{N_R}}$ is given by
\aln{\label{bound_v}
|J_{k,k+1}|\sqrt{\braket{\hat{n}_{k+1}+\hat{n}_k-2\hat{n}_k\hat{n}_{k+1}}^2-\braket{i(\hat{c}_k^\dag\hat{c}_{k+1}-\hat{c}_{k+1}^\dag\hat{c}_k)}^2},
}
which actually depends on the boundary sites $\{k,k+1\}$ and is  useful.
}
{In practice, it is easier to evaluate $\mathcal{\tilde{S}}:=\lrs{\int_0^{t_\mathrm{fin}}dt|J_{k,k+1}|\lrv{\braket{\hat{n}_{k+1}+\hat{n}_k-2\hat{n}_k\hat{n}_{k+1}}}}^2\geq \mathcal{S}$, since the local occupation number in $\mathcal{\tilde{S}}$ is measured by a quantum gas microscope in cold atomic systems~\cite{bakr2009quantum}. Then, we obtain another bound $\mc{\tilde{B}}$ by replacing $\mathcal{S}$ with $\mathcal{\tilde{S}}$ in \eqref{entbound}.}

{
Furthermore, the above bounds for the entanglement entropy can provide tighter bounds than the previous approach~\cite{PhysRevA.76.052319,gong2022bounds} to bound the increase of the  entanglement entropy since the previous bounds are mainly independent of the state $\hat{\rho}$.
In contrast, our bound obtained from \eqref{entbound} and \eqref{bound_v} clearly tells us that the entropy increase is suppressed for certain situations for the state, e.g., when the two boundary sites $\{k,k+1\}$  tend to be simultaneously occupied ($\braket{\hat{n}_k},\braket{\hat{n}_{k+1}},\braket{\hat{n}_k\hat{n}_{k+1}}\simeq 1$) or simultaneously empty ($\braket{\hat{n}_k},\braket{\hat{n}_{k+1}},\braket{\hat{n}_k\hat{n}_{k+1}}\simeq 0$).
}

{
 Figure~\ref{application_quant_m}\textbf{b} shows the time evolution of the entanglement entropy and our {bounds ($\mathcal{B}$ in \eqref{entbound} and $\mathcal{\tilde{B}}$}, as well as the previous general (state-independent) bound based on the entangling rate, $\mathcal{B}':=8(\ln 2)J_{k,k+1}t$~\cite{PhysRevA.76.052319,gong2022bounds}.
{Our bounds provide good results, especially for short-time regime, and that they become much  better than the previous bound.}
}

{
\textbf{Limits from the classical information theory.}
So far, we have considered situations where our results with $c=1$ hold with the condition \eqref{fsit}.  
Now let us discuss another scenario, where Eq.~\eqref{crureq_m} and thus our main results hold for $c=1/2$.
Although the theory applies to both continuous and discrete systems, we here present the case with discrete systems, $\rho=\{p_i\}$.
For an  observable $A=\{A_i\}$, we have
 the average  $(A|\rho)=\braket{A}=\sum_ip_iA_i$ and $(A|\mc{L}[\rho])=\sum_i p_i\delta A_i \dot{p}_i/p_i=(\mc{L}^\dag|\rho)$, where we have defined 
$(\mc{L}^\dag[A])_i=\delta A_i \dot{p}_i/p_i=-\delta A_i\dot{I}_i$, with ${I}_i=-\ln p_i$ called the surprisal~\cite{nicholson2020time}.
Then, $\fracd{\braket{A}}{t}=\braket{\mc{V}_A}$ with Eq.~\eqref{velans} leads to  the continuous-time Price equation~\cite{price1972extension}, $\fracd{\braket{A}}{t}-\dot{A}=-\braket{\delta A \:\dot{I}}=\mr{cov}(A,-\dot{I})$, where  $\mr{cov}(X,Y)=\braket{XY}-\braket{X}\braket{Y}$ is the covariance.
Using the Cauchy-Schwarz inequality, we find the speed limit for the mean~\cite{nicholson2020time}, $\lrv{\fracd{\braket{A}}{t}-\dot{A}}\leq \sigma_A\sqrt{\mc{F}_\mr{C}}$, where $\mc{F}_\mr{C}=\sigma_{\dot{I}}=\sum_i\frac{\dot{p}_i^2}{p_i}$ is the classical Fisher information.
} 

{
In contrast, if we consider the fluctuation, we cannot necessarily obtain our main results \eqref{fluc_main}-\eqref{vel_eqm} for $\mc{V}_A=\dot{A}+\mc{L}^\dag[A]$ since Eq.~\eqref{crureq_m} does not hold.
For example, $\mc{L}^\dag$ does not satisfy Eq.~\eqref{fsit} in general.
To proceed, we focus on the time-independent observable $\dot{A}=0$ in the following.
Then, while Eq.~\eqref{fsit} is not still satisfied, we find that 
Eq.~\eqref{crureq_m} with $c=1/2$ holds for this case, by noticing that $\braket{\braket{A^2}\dot{I}}=\braket{\braket{A}\dot{I}}=0$.
Therefore, for time-independent observables $A$, we have $\fracd{\braket{\delta A^2}}{t}=\braket{\delta A\:\delta \mc{V}_A}=\mr{cov}(A,\mc{V}_A)$, $\lrv{\fracd{\sigma_A}{t}}\leq \frac{\sigma_{\mathcal{V}_A}}{2}$, and 
\aln{\label{infotheo}
\lrs{\fracd{\braket{A}}{t}}^2+4\lrs{\fracd{\sigma_A}{t}}^2\leq C_A
}
with 
\aln{\label{sfit}
(\mc{V}_A)_i=(\mc{L}^\dag[A])_i=-\delta A_i\dot{I}_i. 
}
}

Now, using the H\"older inequality {(instead of the Cauchy-Schwarz inequality as done for the mean~\cite{nicholson2020time,PhysRevX.10.021056}), $C_A=\braket{\mc{V}_A^2}$} is upper bounded as 
\aln{\label{class}
C_A\leq \|\delta A\|_\infty^2\mc{F}_\mr{C}.
}
{Here, $\|\delta A\|_\infty=\max_i |\delta A_i|\leq \Delta_A$, where $\Delta_A$ is the difference between the maximum and minimum values of $A$ and often easily computed.}
Our inequalities mean that, as in the conventional speed limit only for the mean~\cite{nicholson2020time,PhysRevX.10.021056}, the sum of the squared standard deviation and the mean obeys the time-information uncertainty relation, i.e., the times for the change of those fundamental quantities and {the Fisher information (the metric of information geometry)} cannot be simultaneously small.
{
However, we again stress that these results for the fluctuation cannot be inferred from previous results for the mean~\cite{nicholson2020time,PhysRevX.10.021056}, which are obtained without the nontrivial condition in Eq.~\eqref{crureq_m} [see also \textbf{Methods} for the distinctions from \cite{nicholson2020time,PhysRevX.10.021056,PhysRevX.12.011038}].
}

\textit{Example 5: Nonlinear population dynamics.}
Let us consider nonlinear population dynamics, where the number $N_i$ of some types (such as species in an ecological system) labeled by $i=1,\cdots,f$ obeys some  dynamical equation, $\dot{N}_i=F_i(N_1,\cdots, N_f,t)$. Note that $F$ is nonlinear in general.
While the total number of the types $N_\mr{tot}=\sum_{i=1}^fN_i$ changes in time, the proportion for each type $i$, i.e., $p_i=N_i/N_\mr{tot}$, satisfies $\sum_{i=1}^fp_i=1$.
We can then apply our general discussion above.
We stress that our fluctuation relations $\fracd{\braket{\delta A^2}}{t}=\mr{cov}(A,\mc{V}_A)$ and $\lrv{\fracd{\sigma_A}{t}}\leq\frac{\sigma_{\mc{V}_A}}{2}$ are fundamentally distinct from the Price equation, $\fracd{\braket{A}}{t}=\braket{\mc{V}_A}=-\mr{cov}(A,\dot{I})${~\cite{price1972extension} and the speed limit of the mean~\cite{adachi2022universal,garcia2024limits,hoshino2023geometric}.}

We especially take $F_i=s_iN_i$, which describes the evolutionary dynamics without mutation, where $s_i$ is the growth rate for type $i$.
The dynamical equation for $p_i$ then reads $\dot{p}_i=\delta s_i\:p_i$, where $\delta s_i=s_i-\braket{s}$.
In this case, we can obtain $(\mc{V}_A)_i=\delta A_i\delta s_i$
and  $\fracd{\braket{\delta A^2}}{t}=\braket{\delta A^2\delta s}$, which leads to $\lrs{\fracd{\braket{A}}{t}}^2+4\lrs{\fracd{\sigma_A}{t}}^2\leq C_A=\braket{{\delta A^2\delta s^2}}\leq \|\delta A\|_\infty^2\sigma_s^2$.
Thus, the $C_A$ is bounded using the variance of the growth rate, $\sigma_s^2$.

As the simplest case, let us consider $A=s$, for which we have $\fracd{\braket{{s}}}{t}=\braket{\mc{V}_s}=\sigma_s^2$, which is known as the Fisher's fundamental theorem of natural selection~\cite{fisher1999genetical}. In this case, we furthermore find a nontrivial higher-order equality $\fracd{\braket{\delta s^2}}{t}=\braket{\delta s^3}$, meaning that the change of the  variance of the growth rate is related to that of the skewness.

\textbf{Limits in arbitrary quantum dynamics.}
Finally, let us discuss arbitrary quantum dynamics, which may include dissipation and thus non-unitarity.
Because of the non-commutative nature of quantum theory, 
{
we have not found a suitable velocity observable for which \eqref{fluc_main}-\eqref{vel_eqm} hold. 
For example, since the general dynamics is given by $\fracd{\hat{\rho}}{t}=\frac{1}{2}(\hat{\rho}\hat{L}+\hat{L}\hat{\rho})$ using the  symmetric-logarithmic derivative $\hat{L}$~\cite{PhysRevX.12.011038}, it is natural to take $\mc{\hat{V}}_A=\mc{L}^\dag[\hat{A}]+\dot{\hat{A}}$ with
$\mc{L}^\dag[\hat{A}]=\frac{1}{2}\{\hat{A},\hat{L}\}$.
Indeed, this leads to the quantum speed limit for the mean in Ref.~\cite{PhysRevX.12.011038} with the help of the  Cauchy-Schwarz inequality.
However, the above $\mc{\hat{V}}_A$ does \textit{not}  satisfy Eq.~\eqref{crureq_m} (even when $\dot{\hat{A}}=0$), and thus \eqref{fluc_main}-\eqref{vel_eqm} cannot be proved. For another example, if we assume the Lindlad equation for the dynamics, it is natural to expect $\mc{L}^\dag[\hat{A}]=i[\hat{H},\hat{A}]+\sum_\mu \hat{l}_\mu^\dag\hat{A}\hat{l}_\mu-\frac{1}{2}\{\hat{A},\hat{l}_\mu^\dag\hat{l}_\mu\}$, where $\hat{l}_\mu$ is the Lindblad jump operator. However, this does not satisfy Eq.~\eqref{crureq_m}, either.
}

{While our results, e.g., Eq.~\eqref{fluc_main}, may break down for general quantum systems, 
 we can still find an information-theoretical bound for the tradeoff relation} between the changes of the mean and the standard deviation for time-independent observables (see Supplementary Note 4):
\aln{\label{quantumfisher}
\lrs{\fracd{\braket{{\hat{A}}}}{t}}^2+4\lrs{\fracd{\sigma_A}{t}}^2\leq \braket{\hat{L}\delta\hat{A}^2\hat{L}}\leq \|\delta \hat{A}\|_\infty^2 \mc{F}_\mr{Q},
}
where
 $\mc{F}_\mr{Q}=\braket{\hat{L}^2}$ is the quantum Fisher information~\cite{liu2019quantum}.
We recover inequality~\eqref{class} in the classical limit.
We note that inequality~\eqref{quantumfisher} is fundamentally distinct from the (generalized) quantum Cram\'er-Rao inequality $\lrv{\fracd{\braket{{\hat{A}}}}{t}}\leq \sigma_A\sqrt{\mc{F}_\mr{Q}}$ since \eqref{quantumfisher} tells us the upper bound on  the speed of the fluctuation $\fracd{\sigma_A}{t}$.

{
By considering the unitary dynamics and pure state, \eqref{quantumfisher} becomes
\aln{\label{unitary}
\lrs{\fracd{\braket{\hat{A}}}{t}}^2+4\lrs{\fracd{\sigma_A}{t}}^2\leq \frac{4\braket{\delta \hat{H}\delta\hat{A}^2\delta\hat{H}}}{\hbar^2}\leq \frac{4\|\delta \hat{A}\|_\infty^2 \Delta E^2}{\hbar^2},
}
which is the inequality different from \eqref{unit_dyn}.}
This means that the energy fluctuation can be the upper bound on the change of the standard deviations, similar to the spirit of conventional speed limits in isolated quantum systems.
We stress that our result addresses such an energy-fluctuation-based bound on the fluctuation of observables, not mean.
{We also note that this inequality is true even for mixed states, as shown in Supplementary Note 4.}
\newline

{\textbf{\large{Discussion}}}

We here discuss the ditinctions of our work from previous results.
{
Our previous work~\cite{PRXQuantum.3.020319} also 
 considered speed limits for the fluctuation of observables for a specific situation.
 However, the inequalities~\eqref{fluc_main} and \eqref{cost_main} are tighter and more general.
 For tightness, the previous results in~\cite{PRXQuantum.3.020319} lack the term $\lrs{\fracd{\braket{A}}{t}}^2$ in \eqref{cost_main}. 
Consequently, the results in~\cite{PRXQuantum.3.020319} cannot suggest crucial physical implications found in our present manuscript, e.g., the tradeoff relation between the speeds of the mean and the standard deviation.
 For generality, Ref.~\cite{PRXQuantum.3.020319} only treated bounds based on the classical continuity equation of probability; in contrast, we can treat, e.g., bounds in unitary quantum systems for general observables~\eqref{unit_dyn} and information-theoretical bounds~\eqref{infotheo}, which are not obtained from the method in Ref.~\cite{PRXQuantum.3.020319}.
 We also note that the results in Ref.~\cite{PhysRevLett.131.160202} focused on the variance
concerning different quantum trajectories of noisy systems, while we consider variance concerning the quantum fluctuation of the state.
Thus, the quantities of interest are different (e.g., only the latter includes the non-linearity with respect to the state), and the meanings of the obtained inequalities in Ref.~\cite{PhysRevLett.131.160202} and our present paper are conceptually distinct.
}

{
In addition, some previous works studied quantum speed limits for quantum scrambling and correlation functions, which are related to fluctuations of dynamics~\cite{PhysRevX.9.041017,hornedal2022ultimate,PhysRevA.106.042436,carabba2022quantum,Hornedal2023geometricoperator}.
Our approach of directly considering the fluctuation of observables (e.g., standard deviation)
is beneficial because it can primarily characterize the probability distribution of observables along with the mean. 
Another striking consequence of our approach is the discovery of the concise inequalities~\eqref{fluc_main} and \eqref{cost_main}, which lead to profound implications, e.g., a nontrivial tradeoff relation between the speeds of the mean and fluctuation. 
}
\newline

{\textbf{\large{Conclusion}}}

To conclude, we discover a law in far-from-equilibrium dynamics,
stating that the speed of the fluctuation dynamics is always smaller than the fluctuation of the velocity observable.
This can also be rewritten as the tradeoff relation that the sum of the squares of the speeds of the mean and the standard deviation cannot exceed {some quantity $C_A$, which is further bounded by physical quantities (e.g., entropy production) or information-theoretical quantity (i.e., the Fisher information).}
{We stress that our bounds can indicate the \textit{upper} bound for the change of the fluctuation,
 while the conventional speed limit based on the Cram\'er-Rao bound only leads to the \textit{lower} bound on the fluctuation.}

Our results are extended to cases with multiple parameters $\vec{\lambda}=(\lambda_1,\cdots,\lambda_Z)$, instead of the actual time $t$.
We can show limits for the change of the fluctuation of an observable when these parameters are varied; in this case, the bounds are generalized to matrix inequalities  (see Supplementary Note 5).
As the Cram\'er-Rao inequalities for multiple parameters turn out to be useful in the field of (quantum) metrology, our inequalities
may also be applied to such metrological purposes, as well as non-equilibrium statistical mechanics.

Our fluctuation relations can apply to {continuous-time} classical and unitary quantum systems, deterministic and stochastic systems, and even nonlinear and many-body systems.
{In contrast, we have not known yet whether we can extend our theory to broader classes of dynamics, e.g., general quantum systems (which are partially done as in \eqref{quantumfisher}) and discrete-time dynamics.}
It is {also} a crucial future task to harness these relations to control complicated systems at the level of fluctuations.
\newline

{\textbf{\large{Methods}}}

{
\textbf{Possible freedom of choosing velocity observables.}
Here, we discuss the possible freedom of the velocity observables. If $\mc{V}_A$ in \eqref{velans} is given, we can alternatively consider $\mc{\bar{L}}^\dag[A]=\mc{L}^\dag[A]+Z$ and $\mc{\bar{V}}_A=\mc{V}_A+Z$, where the shift operator $Z$ satisfies $\braket{Z}=\braket{A,Z}=0$. Under this condition, we still find $\fracd{\braket{A}}{t}=\braket{\mc{\bar{V}}_A}$.
Since a straightforward calculation leads to the condition Eq.~\eqref{crureq_m} for $\mc{\bar{V}}_A$ and $\mc{\bar{L}}^\dag[A]$,
our main results \eqref{fluc_main}-\eqref{vel_eqm} still hold by replacing $\mc{V}_A$ with $\mc{\bar{V}}_A$.
Therefore, we have the freedom of shifting the velocity observable with such $Z$.
However, the condition for $Z$, $\braket{A,Z}=0$, means that  the $Z$  should be $A$ dependent and thus non-universal, in contrast with our original results for $Z=0$.
}

{
\textbf{Generalization to higher-order fluctuations.}
Here, we consider higher-order absolute central  moments, 
$\mu_{A}^{(n)}:=\braket{|A-\braket{A}|^n}$, and discuss
\aln{\label{higher-order}
\lrv{\fracd{(\mu_{A}^{(n)})^{\frac{1}{n}}}{t}}\leq (\mu_{\mc{V}_A}^{(n)})^{\frac{1}{n}},
}
which reduces to \eqref{fluc_main} for $n=2$.
For this purpose, we focus on classical dynamics since we do not have a proof for \eqref{higher-order} in the quantum case due to the non-commutativity.
First, for a given function $F(z)$ with the derivative $F'(z)$, we have
\aln{
\fracd{\braket{F(\delta A)}}{t}=\braket{\mc{L}^\dag[F(\delta A)]}+\braket{F'(\delta A)(\dot{A}-\braket{\mc{V}_A})}.
}
Now, let us assume
\aln{\label{highassumption}
\mc{L}^\dag[F(\delta A)]=F'(\delta A)\mc{L}^\dag[\delta A],
}
which turns to become \eqref{fsit} by taking $F(z)=z^2$ for linear $\mc{L}^\dag$.
Then we obtain
\aln{
\fracd{\braket{F(\delta A)}}{t}=\braket{F'(\delta A)\delta\mc{V}_A}.
}
Here, we take $F(z)=|z|^n$; then we find, using the H\"older inequality,}
\aln{
\lrv{\fracd{\braket{|\delta A|^n}}{t}}&\leq n\braket{|\delta A|^{n-1}|\delta \mc{V}_A|}
\leq n\braket{|\delta A|^n}^{\frac{n-1}{n}}\braket{|\delta \mc{V}_A|^n}^{\frac{1}{n}}.
}
This leads to the inequality in \eqref{higher-order} for the higher-order absolute central moments $\mu_A^{(n)}$.

{
Now, the assumption \eqref{highassumption} holds  for a classical continuous system with $\mc{L}^\dag[A]=\nabla A\cdot \mbf{V}$.
Therefore, \eqref{higher-order} is valid with $\mc{V}_A=\nabla A\cdot \mbf{V}+\dot{A}$.
This inequality shows that the rate of the change for the higher-order fluctuations of an observable is smaller than the higher-order fluctuations of the change of the velocity of the observable.
}

\textbf{Relation to the Wasserstein geometry.}
In the context of optimal transport theory, it has been shown that minimization of $\int d\mbf{x}\rho(\mbf{x},t)|\mbf{V}(\mbf{x},t)|^2$ over $\mbf{V}$ satisfying the continuity equation leads to~\cite{PhysRevResearch.3.043093} 
${\lim_{\delta t\ra 0}\frac{\mc{W}_2(\rho(t),\rho(t+\delta t))^2}{\delta t^2}}$, which is regarded as the Benamou-Brenier formula~\cite{benamou2000computational} for infinitesimal times.
Here, $\mc{W}_2(\rho,\rho')$ is the order-2 Wasserstein distance between two densities $\rho$ and $\rho'$ (see, e.g., Ref.~\cite{villani2009optimal} for the definition).
Then, our result indicates
\aln{\label{opt}
&\lrs{\fracd{\braket{{A}}}{t}}^2+\lrs{\fracd{{\sigma}_A}{t}}^2\leq C_A\nonumber\\
&\leq \|\nabla A\|_\infty^2\lrs{\lim_{\delta t\ra 0}\frac{\mc{W}_2(\rho(t),\rho(t+\delta t))^2}{\delta t^2}}.
}

For simplicity, let us consider a one-dimensional system. Then, the inequality proposed in Ref.~\cite{gelbrich1990formula} leads to $(\braket{x(t)}-\braket{x(t+\delta t)})^2+(\sigma_x(t)-\sigma_x(t+\delta t))^2\leq \mc{W}_2(\rho(t),\rho(t+\delta t))^2$.
Thus, taking $\delta t\ra 0$ limit, we obtain \eqref{opt} for $A=x$ since $\|\nabla x\|_\infty=1$.
Note that this inequality achieves the equality condition when $\rho(t)$ and $\rho(t+\delta t)$ are Gaussian.
Thus, for such Gaussian processes, our inequalities in \eqref{cost_main} and $\eqref{cost_ev}$ can satisfy the equality condition with $A=x$.
{More generally, we can also show that the knowledge on the Wasserstein distance leads to \eqref{now2} or \eqref{opt} for a general observable $A$, as detailed in Supplementary Note 6.}

We also note that for irreversible thermodynamics discussed in {\textit{Example 1}}, the right-hand side of \eqref{opt} can be given by $\|\nabla A\|^2\mu T\dot{\Sigma}^\mathrm{ex}$, where $\dot{\Sigma}^\mathrm{ex}\:(\leq \dot{\Sigma})$ is the Maes-Neto\v{c}n\`{y} excess entropy production rate~\cite{maes2014nonequilibrium,PhysRevResearch.4.L012034}. Therefore, $C_A$ is also bounded by the excess entropy production rate in our inequality \eqref{cost_main}, 
which is tighter than the bound by the total entropy production rate.

\textbf{Extension of the current-based bound to discrete systems.}

{
While we have mainly discussed the bounds based on the local conservation law for continuous systems in the main text, a similar structure appears for discrete systems.
To see this, let us consider a graph $G$, which consists of vertices and edges and can model general discrete systems~\cite{PRXQuantum.3.020319}.
Here, the vertices are taken as the basis $i$ of the system, and the edges are connected between vertices if and only if there can be transitions between $i$ and $j$ upon dynamics.
We define the normalized distribution $p_i$ ($\sum_ip_i=1$) and an observable $A=\{A_i\}$ defined on the vertices.
We assume that the discrete continuity equation $\dot{p}_i=-\sum_{j(\sim i)}J_{ji}$ holds, where $J_{ji}\:(=-J_{ij})$ is a current from $i$ to $j$ and $i\sim j$ means that $i$ and $j$ are connected by the edges. 
}

{
Even in this case, we show that Eqs.~\eqref{fluc_main}-\eqref{vel_eqm} still hold, where each of the quantities is properly generalized to the discrete ones.} For this purpose, we first introduce a new probability measure $Q_{ij}$ defined on the edges, which is assumed to satisfy $p_i=\sum_{j(\sim i)}Q_{ij}$ (note that $Q_{ij}\neq Q_{ji}$ in general).
Next, we define the mean $\braket{O}_Q$ (and similarly, the standard deviation $\sigma_O^Q$) of an observable $O_{ij}$ defined on the edges as $\braket{O}_Q=\sum_{i\sim j} O_{ij}Q_{ij}$.
We also consider the extension of an observable $A_i$ defined on the vertices to that for the vertices by defining $A_{ij}=A_i$.
Then, $\braket{A}_Q=\sum_{ij}A_iQ_{ij}=\sum_iA_ip_i=\braket{A}$, and similarly, $\sigma_A^Q=\sigma_A$.

{
Due to the existence of two types of averages, $\braket{\cdots}$ and $\braket{\cdots}_Q$, we cannot directly apply the method in \textbf{General proof and the required condition.}; however, the structure of the proof is essentially the same.
}
We first note that 
\aln{
\fracd{\braket{A}}{t}&=\sum_i(A_i\dot{p}_i+\dot{A}_ip_i)\nonumber\\
&=
\braket{\dot{A}}+\frac{1}{2}\sum_{i\sim j}(A_i-A_j)J_{ij}\nonumber\\
&=\sum_{i\sim j}Q_{ij}\dot{A}_i+\frac{1}{2}\sum_{i\sim j}Q_{ij}(\nabla A)_{ij}V_{ij}=\braket{\mc{V}_A}_Q,
}
where the velocity observable is given by 
\aln{
(\mc{V}_A)_{ij}=\frac{1}{2}(\nabla A)_{ij}V_{ij}+\dot{A}_i
}
with $(\nabla A)_{ij}=A_i-A_j$ and $V_{ij}=J_{ij}/Q_{ij}$. 
Here, we have used $J_{ij}=-J_{ji}$ from the second to the third lines.
{From this calculation, it is also natural to define $\mc{\tilde{L}}^\dag[A]$ as
\aln{\label{Ldagdis}
(\mc{\tilde{L}}^\dag[A])_{ij}=\frac{1}{2}(\nabla A)_{ij}V_{ij},
}
which satisfies $\sum_iA_i\dot{p}_i=\braket{\mc{\tilde{L}}^\dag[A]}_Q$.
If we use the Cauchy-Schwarz inequality as $\lrv{\fracd{\braket{A}}{t}-\dot{A}}=|\braket{\mc{\tilde{L}}^\dag[A]}_Q|\leq \frac{1}{2}\sqrt{\braket{\nabla A^2}}\sqrt{\braket{\mbf{V}^2}}$, 
we obtain the quantum speed limits for the mean of $A$ discussed previously~\cite{PRXQuantum.3.020319,hamazaki2023quantum}. However, it is nontrivial if our main results~\eqref{fluc_main}-\eqref{vel_eqm} on fluctuation dynamics hold.
}

{To discuss the fluctuation dynamics, we notice the following, which is obtained from the straightforward calculation
\aln{
\fracd{\braket{(\delta A)^2}}{t}
&=\braket{2\delta A\:\dot{A}+\mc{\tilde{L}}^\dag[A^2]-2\braket{A}\mc{\tilde{L}}^\dag[A]}_Q
}
and is similar to the left-hand side of Eq.~\eqref{crureq_m}.
Now, using \eqref{Ldagdis}, we find $\braket{\mc{\tilde{L}}^\dag[A^2]}_Q=2\braket{A\mc{\tilde{L}}^\dag[A]}_Q$, which is similar to \eqref{fsit}.
Then, we obtain
\aln{
\fracd{\braket{(\delta A)^2}}{t}=2\braket{\delta A\delta\mc{V}_A}_Q,
}
in a manner similar to Eq.~\eqref{vel_eqm} with $c=1$.
}

Using the Cauchy-Schwarz inequality, we have
$\lrv{\fracd{\braket{(\delta A)^2}}{t}}\leq 2\sqrt{\braket{\delta A^2}_Q\braket{\delta \mc{V}_A^2}_Q}$ and
thus (noting $\sigma_A=\sigma_A^Q$) 
\aln{
\lrv{\fracd{\sigma_A^Q}{t}}\leq \sigma_{\mc{V}_A}^Q
}
and
\aln{
\lrs{\fracd{\braket{A}}{t}}^2+\lrs{\fracd{\sigma_A}{t}}^2\leq C_A=\braket{\mc{V}_A^2}_Q
}
for the discrete case.

{
Furthermore, using the H\"older's inequality (instead of the Cauchy-Schwarz inequality), we find 
\aln{
C_A\leq \frac{1}{4}\|\nabla A\|_\infty^2 \braket{\mbf{V}^2},
}
where $\|\nabla A\|_\infty=\max_{i\sim j}|A_i-A_j|$ is the magnitude of the discrete gradient.
}

We can apply our general theory to quantum transport in macroscopic systems:

{
\textit{Example 6: Macroscopic transport in quantum many-body systems.}}
Let us consider macroscopic transport in quantum many-body systems.
We consider a many-body basis set $\{\ket{i}\}$ as a set of vertices for $G$ and focus on a time-independent observable $\hat{A}=\sum_iA_i\ket{i}\bra{i}$, which is diagonalized in this basis. 
For example, we can take $\hat{A}$ as the averaged position of $M$ particles on a one-dimensional system, $\hat{A}=\frac{1}{M}\sum_ll\hat{n}_l$, where $\hat{n}_l$ is the number operator for site $l$; then the changes of mean and fluctuation of $\hat{A}$ correspond to how macroscopic transport of particles takes place.
Now, the expectation value is given by $\braket{\hat{A}}=\sum_iA_ip_i$, where $p_i=\braket{i|\hat{\rho}|i}$ is the probability distribution. Here, $p_i$ satisfies the continuity equation $\dot{p}_i=-\sum_{j(\sim i)}J^q_{ij}$, where $J^q_{ij}=-i(\braket{i|\hat{H}|j}\braket{j|\hat{\rho}|i}-\mr{h.c.})/\hbar$ and $i\sim j$ if and only if $\braket{i|\hat{H}|j}\neq 0$.
In this case,
we find inequality~\eqref{cost_main} with
the bound on $C_A$ as
$C_A\leq \|\nabla A\|_\infty^2({\mathcal{S}_H^2-E_\mr{kin}^2})/\hbar^2$, where 
$\mathcal{S}_H=\max_j\sum_{i(\sim j)}|\braket{i|\hat{H}|j}|$ is the strength of the transition, which is easily known from the Hamiltonian (see Supplementary Note 1).
The transition part of the energy, $E_\mr{kin}=\braket{\hat{H}}-\sum_ip_i\braket{i|\hat{H}|i}$, is a globally defined macroscopic observable, which is relevant for experiments.
Our inequality $\lrs{\fracd{{\braket{\hat{A}}}}{t}}^2+\lrs{\fracd{{\sigma_A}}{t}}^2\leq C_A\leq \|\nabla A\|_\infty^2({\mathcal{S}_H^2-E_\mr{kin}^2})/\hbar^2$ is tighter than those found in Ref.~\cite{PRXQuantum.3.020319,PhysRevLett.130.010402}. 
{For example, Ref.~\cite{PRXQuantum.3.020319} only derived $\lrs{\fracd{{\sigma_A}}{t}}^2\leq \|\nabla A\|_\infty^2({\mathcal{S}_H^2-E_\mr{kin}^2})/\hbar^2$.}

{\textbf{Semi-classical limit and limits to
classical Hamiltonian dynamics.}}

{
Importantly, our bound for unitary quantum dynamics in \eqref{unit_dyn} remains useful in the semi-classical limit, in stark contrast with the Mandelstam-Tamm bound~\cite{mandelstam1945energy}, which becomes meaningless for $\hbar\ra 0$.
For simplicity, let us focus on the system parametrized by the canonical variables $\hat{q}_k,\hat{p}_k\:(1\leq k \leq f)$.
We then find $\lrv{\fracd{\sigma_A}{t}}\leq \sigma_{\{A,H\}_\mr{Po}}$ and $\lrs{\fracd{\braket{A}}{t}}^2+\lrs{\fracd{\sigma_A}{t}}^2\leq \braket{\{A,H\}_\mr{Po}^2}$, by taking the semi-classical limit $\hbar\rightarrow 0$ in Eq.~\eqref{unit_dyn}.
Here, the average is concerning the Wigner phase-space distribution, $A=A(\vec{q},\vec{p})$ (which we assume to be time-independent for simplicity) and $H=H(\vec{q},\vec{p})$ are the semi-classical limit of the observable and the Hamiltonian, respectively, and $\{X,Y\}_\mr{Po}=\sum_{k=1}^f\partial_{q_k}X\partial_{p_k}Y-\partial_{q_k}Y\partial_{p_k}X$ is the Poisson bracket.
}

{
In the following, we show another derivation of these inequalities from the classical setting and the higher-order generalizations as in \eqref{higher-order}.
}
We denote the classical phase-space probability distribution by $\rho=\{f(\vec{q},\vec{p},t)\}$, which obeys the Liouville's equation $\fracpd{f}{t}=\{H,f\}_\mr{Po}$.
The expectation value of a general time-dependent observable is given by $\braket{A(t)}=\int d\Gamma A(\vec{q},\vec{p},t)f(\vec{q},\vec{p},t)$, where $d\Gamma$ is the phase-space volume element.
Then, using $\int d\Gamma X \{Z,Y\}_\mr{Po}=-\int d\Gamma \{Z,X \}_\mr{Po}Y$, {we find
$(A|\mc{L}[\rho])=(\mc{L}^\dag[A]|\rho)$ with $\mc{L}^\dag[A]=\{A,H\}_\mr{Po}$.
Then, using \eqref{velans}, we can choose the velocity observable as
\aln{
\mc{V}_A=\{A,H\}_\mr{Po}+\dot{A}.
}
}

{
Now, for the above $\mc{L}^\dag[A]$, we can confirm that \eqref{fsit} holds true.
Thus, we obtain $\fracd{\braket{\delta A^2}}{t}=2\braket{\delta A\delta \mc{V}_A}$, $\lrv{\fracd{\sigma_A}{t}}\leq \sigma_{\{A,H\}_\mr{Po}}$, and $\lrs{\fracd{\braket{A}}{t}}^2+\lrs{\fracd{\sigma_A}{t}}^2\leq \braket{\{A,H\}_\mr{Po}^2}$ for time-independent $A$.
Furthermore, we can also verify Eq.~\eqref{highassumption}, which leads to the higher-order version of our inequality, \eqref{higher-order}.
}

Let us show the two simplest examples for the Hamiltonian written as $H=\sum_k\frac{{p}_k^2}{2m_k}+V(\vec{q})$.
For $A=q_k$, we have $\mc{V}_A=\frac{p_k}{m_k}$, which corresponds to $\fracd{\braket{q_k}}{t}=\braketL{\frac{p_k}{m_k}}$.
Importantly, our results indicate a nontrivial fluctuation identity
\aln{
\fracd{\braket{(\delta q_k)^2}}{t}=\frac{2\braket{\delta q_k\delta p_k}}{m_k},
}
i.e., the variance of the position is determined by the covariance between the position and momentum. This also indicates
\aln{
\lrv{\fracd{\sigma_{q_k}}{t}}\leq  \frac{\sigma_{p_k}}{m_k}
}
and
\aln{
\lrs{\fracd{\braket{q_k}}{t}}^2+\lrs{\fracd{\sigma_{q_k}}{t}}^2
\leq \frac{\braket{p_k^2}}{m_k^2}=\frac{2\epsilon_k}{m_k},
}
where $\epsilon_k=\frac{\braket{p_k^2}}{2m_k}$ is the mean value of the kinetic energy for the $k$th component.

As the other example, we take $A=p_k$ and have $\mc{V}_A=-\fracpd{V}{q_k}:=\mathsf{F}_k$, which corresponds to $\fracd{\braket{p_k}}{t}=-\braketL{\fracpd{V}{q_k}}=\braket{\mathsf{F}_k}$, where $\mathsf{F}_k$ is the force for the $k$th component.
In this case, we have
\aln{
\fracd{\braket{(\delta p_k)^2}}{t}&={2\braket{\delta p_k\delta \mathsf{F}_k}},\\
\lrv{\fracd{\sigma_{p_k}}{t}}&\leq  {\sigma_{\mathsf{F}_k}},
}
and
\aln{
\lrs{\fracd{\braket{p_k}}{t}}^2+\lrs{\fracd{\sigma_{p_k}}{t}}^2
\leq {\braket{\mathsf{F}_k^2}}.
}

{\textbf{Bounds on the entanglement entropy.}
If we assume a Gaussian state with no initial entanglement, for which $S_R(0)=\sigma_{N_R}(0)=0$, 
inequalities \eqref{entineq} and \eqref{flucint}  lead to 
\aln{
S_R\leq \mc{S}\ln y +\frac{4|R|}{ey}
}
for $y\geq 1$.
Taking the derivative with $y$, we find that the right-hand side takes the minimum for $y_c=\frac{4|R|}{e\mc{S}}$ as long as $y_c\geq 1$.
Then, we have
\aln{
S_R\leq \mc{S}\ln\frac{4|R|}{\mc{S}}
}
for $\mc{S}\leq \frac{4|R|}{e}$.
At the same time, we have a trivial bound for the entanglement entropy, $S_R\leq |R|\ln 2$.
Thus, for $\mc{S}\geq \frac{|R|}{4}$, $S_R\leq |R|\ln 2$ becomes better.
In summary, we have
\aln{
S_{R}(t_\mathrm{fin})\leq \mc{B}:=
\left\{
\begin{array}{ll}
\mc{S}\:{\ln\frac{4|R|}{\mc{S}}} & \quad (\mc{S}\leq \frac{|R|}{4}),\\
{|R|\ln 2} &\quad (\mc{S}\geq \frac{|R|}{4}).
\end{array}
\right.
}
}

{
Next, we discuss that $\mc{\hat{V}}_R$ (and thus $\sigma_{\mc{V}_{N_R}}$, $\mc{S}$, and $\mc{B}$) can be expressed using the boundary terms of $R$ alone, which is practically useful.
Here, the boundary of $R$ is defined from a set of sites $(i,j)$, which satisfy $J_{ij}\neq 0$, $j\in R$ and $i\notin R$.
Indeed, using $\hat{H}=\sum_{i,j}J_{ij}(\hat{c}_i^\dag \hat{c}_j +\mathrm{h.c.})+\sum_i V_i\hat{n}_i$ and $\hat{N}_R=\sum_{i\in R}\hat{n}_i$, we obtain
\aln{
&\mc{\hat{V}}_R=i[\hat{H},\hat{N}_R]\nonumber\\
&=i\sum_{i,j}\sum_{k\in R}J_{ij}(\delta_{jk}\hat{c}_i^\dag \hat{c}_k-\delta_{ik}\hat{c}_k^\dag \hat{c}_j)\nonumber\\
&=i\sum_{i,j}\sum_{k\in R}J_{ij}(\delta_{jk}\hat{c}_i^\dag \hat{c}_k-\delta_{jk}\hat{c}_k^\dag \hat{c}_i)\nonumber\\
&=i\sum_i\sum_{j\in R}J_{ij}(\hat{c}_i^\dag \hat{c}_j-\hat{c}_j^\dag \hat{c}_i)=i\sum_{i\notin \bar{R}}\sum_{j\in R}J_{ij}(\hat{c}_i^\dag \hat{c}_j-\hat{c}_j^\dag \hat{c}_i),
}
which only acts on the boundary of $R$ (we have used $J_{ij}=J_{ji}$).
}

{\textbf{Distinction between Eq.~\eqref{vel_eqm} for the information-theoretical context and the results in Refs.~\cite{nicholson2020time,PhysRevX.10.021056,PhysRevX.12.011038}.}
As  discussed in the main text, the results in Refs.~\cite{nicholson2020time,PhysRevX.10.021056} are based on the continuous-time Price equation, which reads
\aln{\label{Pricemeth}
\fracd{\braket{A}}{t}=\braket{\delta A(-\dot{I})}=\braket{\delta A\delta (-\dot{I})}=\mr{cov}(A,-\dot{I})=\braket{\mc{V}_A},
}
where $\mc{V}_A$ is given in \eqref{sfit}.
Here, we have assumed that the observables are time-independent.
Therefore, the Price equation just provides the definition for the velocity observable.
By using the Cauchy-Schwarz inequality and $\braket{\dot{I}^2}=\mc{F}_\mr{C}$, we find the speed limit for the mean of $A$, $\lrv{\fracd{\braket{A}}{t}}\leq \sigma_A\sqrt{\mc{F}_\mr{C}}$.}

{
In contrast, our limits to fluctuation dynamics for the above $\mc{V}_A$ are obtained by considering Eq.~\eqref{vel_eqm} with $c=1/2$, i.e.,
\aln{\label{CVEm}
\fracd{\braket{\delta{A}^2}}{t}=\braket{\delta A\delta\mc{V}_A}=\mr{cov}(A,\mc{V}_A),
}
which has not been obtained before.
We stress that this cannot be trivially obtained from Eq.~\eqref{Pricemeth}.
Indeed, while it seems that \eqref{CVEm} would be obtained by substituting $A$ in Eq.~\eqref{Pricemeth} with $F(\delta A)\delta A$ ($F$ is some function) as 
\aln{
\fracd{\braket{F(\delta A)\delta{A}}}{t}&\overset{?}{=}\braket{F(\delta A)\delta A\delta(-\dot{I})}\nonumber\\
&=\braket{F(\delta A)\delta \mc{V}_A}=\mr{cov}(F(\delta A),\mc{V}_A)
}
and taking $F(\delta A)=\delta A$, the first equation is incorrect in general because of the non-trivial dependence  of $F(\delta A)$ on $t$ and $\{p_i\}$.
Note that while the equation happens to be correct for $F(\delta A)=\delta A$, this is a coincidence, where we carefully need to check the condition~\eqref{crureq_m}. 
For example, a naive extension to the higher-order fluctuation cannot hold for this case, $\fracd{\braket{\delta{A}^3}}{t}\neq\mr{cov}(\delta A^3,-\dot{I})=\mr{cov}(\delta A^2,\mc{V}_A)$.
Moreover, the fact that \eqref{CVEm} does not trivially follow from \eqref{Pricemeth} is known from their quantum generalization; while quantum extension of Eq.~\eqref{Pricemeth} holds as discussed in Ref.~\cite{PhysRevX.12.011038}, the naive quantum extension in \eqref{CVEm} fails for general open quantum systems.
}

\:
\newline

{\textbf{\large{Data availability}}}

{
All the data that support the plots and other findings of this study are available from the corresponding author upon reasonable request.}
\newline

{\textbf{\large{Code availability}}}

{
All the computational codes that were used to generate the data presented in this study are available from the corresponding authors upon reasonable request.
The numerical calculations were carried out with the help of QUSPIN (0.3.6)~\cite{weinberg2017quspin,weinberg2019quspin}.
}
\newline

\bibliography{reference}

\ 
\newline

{\textbf{\large{Acknowledgments}}}

We thank Kyosuke Adachi, Takashi Mori, {Yuichiro Matsuzaki, and Ryosuke Iritani} for fruitful comments.
 { This work was supported by JST ERATO Grant No. JPMJER2302, Japan, and by JSPS KAKENHI Grant No. JP24K16982.}
\newline

{\textbf{\large{Author Contributions}}}

This work was done by R.H.

\ 
\newline

{\textbf{\large{Competing financial interests}}}

The authors declare no competing interests.

\end{document}


\title{
 Supplementary Information for ``{Speed limits} to fluctuation dynamics"
}
\author[1]{Ryusuke Hamazaki}
\affil[1]{Nonequilibrium Quantum Statistical Mechanics RIKEN Hakubi Research Team, RIKEN Cluster for Pioneering Research (CPR), RIKEN iTHEMS, Wako, Saitama 351-0198, Japan}



\renewcommand{\thetable}{Supplementary Equation \arabic{table}}

\renewcommand{\refname}{Supplementary References}

\maketitle

\section*{Supplementary Note 1: 
Proof of the bound for macroscopic quantum transport in {\textit{Example 6}} in the main text.}

Here, we apply the general discussion presented in ``\textbf{Method: Extension of the current-based bound to discrete systems}" to the unitary quantum dynamics of a time-independent observable $\hat{A}=\sum_iA_i\ket{i}\bra{i}$ ({\textit{Example 6}} in the main text).
For simplicity, we set $\hbar=1$ in this Supplementary Information.
In this case, the discrete quantum current reads
\aln{
J_{ij}^q=-i(H_{ij}\rho_{ji}-\rho_{ij}H_{ji}),
}
where $H_{ij}=\braket{i|\hat{H}|j}$ and $\rho_{ij}=\braket{i|\hat{\rho}|j}$.
We take 
\aln{
Q_{ij}=\frac{p_i|H_{ij}|}{\sum_{k(\sim i)}|H_{ik}|},
}
which satisfies $p_i=\sum_{j(\sim i)}Q_{ij}$.

From the general discussion, we find
\aln{
C_A=
\sum_{i\sim j}\frac{(A_i-A_j)^2(J_{ij}^q)^2}{4Q_{ij}}.
}
Now, let us show that the $C_A$ is upper bounded using the modified transition strength $\mc{S}_H^2-E_\mr{trans}^2$, where
\aln{
\mc{S}_H=\max_i\sum_{j(\sim i)}|H_{ij}|
}
is the strength of the transition and 
\aln{
E_\mr{trans}=\sum_{i\sim j}H_{ij}\rho_{ji}=\braket{\hat{H}}-\sum_ip_iH_{ii}
}
is the transition part of the energy (note that $\sum_ip_iH_{ii}$ provides the sum of the on-site potential and the particle interactions for systems of particles).

We have
\aln{
C_A&\leq \frac{\|\nabla A\|_\infty^2}{4}\sum_{i\sim j}\frac{(J_{ij}^q)^2}{Q_{ij}}\nonumber\\
&\leq\frac{\|\nabla A\|_\infty^2}{4}\sum_{j}\sum_{i(\sim j)}\frac{(J_{ij}^q)^2}{p_i|H_{ij}|}\sum_{k(\sim i)}|H_{ik}|\nonumber\\
&\leq
\frac{\|\nabla A\|_\infty^2}{4}\sum_{j}\lrs{\max_{i(\sim j)}\sum_{k(\sim i)}|H_{ik}|}\sum_{i(\sim j)}\frac{(J_{ij}^q)^2}{p_i|H_{ij}|}\nonumber\\
&\leq
\frac{\|\nabla A\|_\infty^2\mc{S}_H}{4}\sum_{i\sim j}\frac{(J_{ij}^q)^2}{p_i|H_{ij}|}\nonumber\\
}
Now, as shown in Ref.~\cite{PRXQuantum.3.020319} (especially see page 29), we can show 
\aln{
\sum_{i\sim j}\frac{(J_{ij}^q)^2}{p_i|H_{ij}|}
\leq 4\mc{S}_H-\frac{4E_\mr{trans}^2}{\mc{S}_H}.
}
Thus, we finally have
\aln{\label{supptra}
\lrs{\fracd{\braket{{\hat{A}}}}{t}}^2+\lrs{\fracd{\sigma_A}{t}}^2\leq C_A\leq \|\nabla A\|_\infty^2(\mc{S}_H^2-E_\mr{trans}^2).
}
This means that $C_A$ becomes smaller if the transition energy increases.

{
We note that our previous work in Ref.~\cite{PRXQuantum.3.020319} obtained
\aln{\label{supppre}
\lrv{\fracd{\sigma_A}{t}}\leq  \|\nabla A\|_\infty\sqrt{\mc{S}_H^2-E_\mr{trans}^2}.
}
Therefore, our present inequality \eqref{supptra} becomes stronger than \eqref{supppre}.
}

\section*{Supplementary Note 2: Numerical demonstration for quantum many-body systems.}
Here, we numerically demonstrate our limit to fluctuation dynamics,
\aln{\label{unitaryfluc}
\lrv{\fracd{\sigma_{A}}{t}}\leq \sigma_{i[\hat{H},\hat{A}]},
}
or, equivalently, 
\aln{\label{unitarymt}
\lrs{\fracd{\braket{\hat{A}}}{t}}^2+\lrs{\fracd{\sigma_{A}}{t}}^2\leq -\braket{[\hat{H},\hat{A}]^2},
}
for concrete quantum many-body systems.

\begin{figure}
\begin{center}
\includegraphics[width=\linewidth]{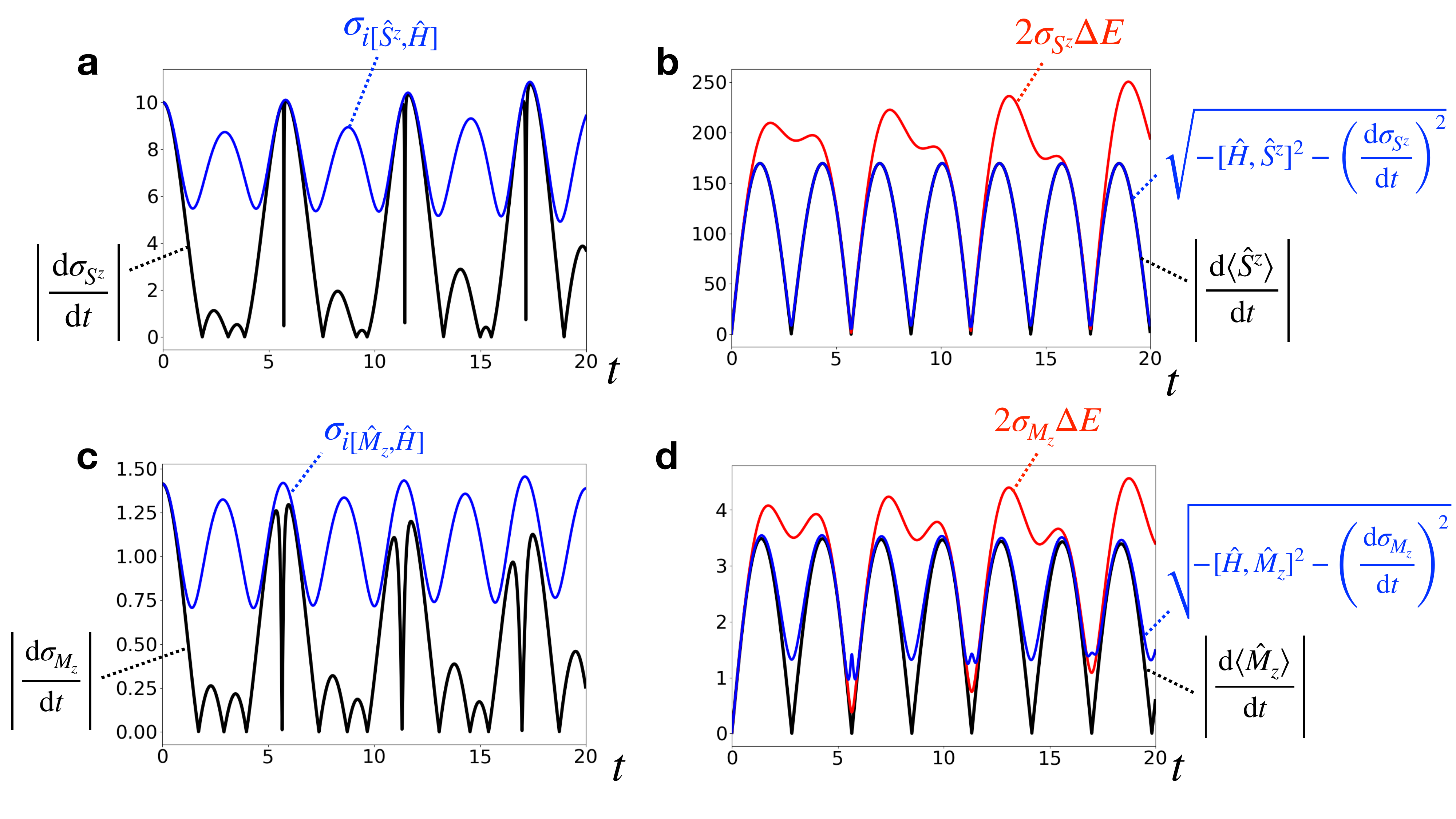}
\end{center}
\caption{Supplementary Figure S-1.
\textbf{Time evolution of the rate of the change for the fluctuation and expectation value, as well as their bounds.}
\textbf{a} Time evolution of $\lrv{\fracd{\sigma_{S^z}}{t}}$ (black) in the model in Eq.~\eqref{alltoall}. Our bound in \eqref{unitaryfluc} (blue) indeed hold and can be quite tight in some time intervals.
\textbf{b} Time evolution of $\lrv{\fracd{\braket{\hat{S}^z}}{t}}$ (black) in the model in Eq.~\eqref{alltoall}.
Our bound $\sqrt{-\lrs{\fracd{\sigma_{S^z}}{t}}^2-\braket{[\hat{H},\hat{S}^z]^2}}$ (blue) can be much tighter than the Mandelstam-Tamm speed limit $\sigma_{S^z}\Delta E$ (red). Note that black and blue lines are almost overlapped. We take $S=200$.
\textbf{c} Time evolution of $\lrv{\fracd{\sigma_{M_z}}{t}}$ (black) with $\hat{M}_z=\sum_{i=1}^L\hat{s}^z_i$ in the model in Eq.~\eqref{local}. Our bound in \eqref{unitaryfluc} (blue) indeed hold and can be quite tight in some time intervals.
\textbf{d} Time evolution of $\lrv{\fracd{\braket{\hat{M}_z}}{t}}$ (black) in the model in Eq.~\eqref{local}.
Our bound $\sqrt{-\lrs{\fracd{\sigma_{{M}_z}}{t}}^2-\braket{[\hat{H},\hat{M}_z]^2}}$ (blue) can be much tighter than the Mandelstam-Tamm speed limit $\sigma_{M_z}\Delta E$ (red). We take $L=8$.
For all of the cases, we use $J=0.1,g=1.0$, and $h=0.5$.
The initial state is  the state where all spins point $+z$ direction.
}
\label{nummany}
\end{figure}

The first model we consider is a system with all-to-all couplings, whose Hamiltonian is given by
\aln{\label{alltoall}
\hat{H}=\frac{J}{S}(\hat{S}^z)^2+g\hat{S}^x+h\hat{S}^z,
}
where $\hat{S}^{x,y,z}$ is the spin-$S$ spin operator.
Figure~\ref{nummany}\textbf{a} shows the rate of the change of the fluctuation for $\hat{A}=\hat{S}^z$ and its bound based on \eqref{unitaryfluc} for $S=200$. We can see that our bound can be quite tight in some time intervals.
Furthermore, Fig.~\ref{nummany}\textbf{b} shows the speed of $\braket{\hat{A}}$ and its bounds defined from the bound based on \eqref{unitarymt} and from the Mandelstam-Tamm quantum speed limit, $\lrv{\fracd{\braket{\hat{A}}}{t}}\leq 2\sigma_A\Delta E$.
Notably, we can see that the bound based on \eqref{unitarymt} can be tighter than the Mandelstam-Tamm quantum speed limit.

The second model is a one-dimensional locally interacting quantum many-body system given by
\aln{\label{local}
\hat{H}=\sum_{i=1}^LJ\hat{s}_i^z\hat{s}_{i+1}^z+g\hat{s}_i^x+h\hat{s}_i^z,
}
where  $\hat{s}_i^{x,y,z}$ is the spin-1/2 spin operator at site $i$, and  we impose the periodic boundary condition.
Figures~\ref{nummany}\textbf{c} and \textbf{d} respectively show $\lrv{\fracd{\sigma_A}{t}}$ and $\lrv{\fracd{\braket{\hat{A}}}{t}}$ for $\hat{A}=\hat{M}_z=\sum_i\hat{s}_i^z$ and their bounds as in Figs.~\ref{nummany}\textbf{a} and \textbf{b}.
We again find that our bounds work well and can be tighter than the Mandelstam-Tamm quantum speed limit.

{
\section*{Supplementary Note 3: Proof of Eq.~(15) in the main text.}
Here, we show the proof of Eq.~(15) in the main text by developing the approach in Ref.~\cite{PhysRevLett.127.090601}.
For Gaussian states of fermions, the reduced density matrix in $R$ is completely characterized by the $|R|\times|R|$ correlation matrix $\Gamma$, whose elements are given by $\Gamma_{ij}=\braket{c_i^\dag c_j}$ with $i\in R, j\in R$. 
Then, the von Neumann entanglement entropy and the number fluctuation are respectively given as
\aln{
S_R=-\sum_{\alpha=1}^{|R|}\lambda_\alpha\ln\lambda_\alpha-(1-\lambda_\alpha)\ln(1-\lambda_\alpha)
}
and
\aln{
\sigma_{N_R}^2=\sum_{\alpha=1}^{|R|}\lambda_\alpha(1-\lambda_\alpha),
}
where $\lambda_\alpha\in [0,1]$ are the eigenvalues of $\Gamma$ (we consider $0\ln 0=0$).
}

{
Now, we can show
\aln{\label{xlog}
-x\ln x-(1-x)\ln(1-x)\leq x(1-x)\ln y +\frac{4}{ey}
}
for any $x\in[0,1]$ and $y>0$.
Note that this is stronger and much more general compared with the result obtained in Ref.~\cite{PhysRevLett.127.090601}.
Since the inequality is obvious for $x=0$ or $x=1$, we consider $x\in(0,1)$ in the following.
}

{
To show \eqref{xlog}, we consider the minimum of
\aln{
G(x,y)=x(1-x)\ln y +\frac{4}{ey}+x\ln x+(1-x)\ln(1-x).
}
Since $\partial_yG(x,y)=x(1-x)/y-4/ey^2$, $G$ takes a minimum value at $y=\frac{4}{ex(1-x)}$ for a fixed $x$.
Then we should show
\aln{
\tilde{G}(x)&=x(1-x)\ln \frac{4}{ex(1-x)}+x(1-x)+x\ln x+(1-x)\ln(1-x)\nonumber\\
&=x(1-x)\ln 4+x^2\ln x+(1-x)^2\ln(1-x)
\geq 0.
}
Since $\tilde{G}(x)$ is symmetric with respect to $x=1/2$, we can focus on $x\in (0,1/2]$ without loss of generality.}

{
To show the non-negativity of $\tilde{G}$, we notice $\tilde{G}'(x)=(1-2x)\ln 4+2x\ln x+2(x-1)\ln(1-x)+2x-1$,
 $\tilde{G}''(x)=-2\ln 4 +6 +2\ln x(1-x)$, and $\tilde{G}'''(x)=2/x-2/(1-x)\geq 0\:\:(x\in(0,1/2])$.
 Then, $\tilde{G}''(x)$ is a monotonically increasing function for $x\in(0,1/2]$. Moreover, since $\tilde{G}''(x\ra 0)= -\infty$ and $\tilde{G}''(1/2)=6-4\ln 4>0$, we find unique $x_c^{(2)}\in (0,1/2]$ such that $\tilde{G}''(x)<0$ for $x<x_c^{(2)}$, $\tilde{G}''(x)=0$ for $x=x_c^{(2)}$, and $\tilde{G}''(x)>0$ for $x>x_c^{(2)}$.
Thus, $\tilde{G}'(x)$ is monotonically decreasing for $0<x<x_c^{(2)}$ and increasing for $x_c^{(2)}<x\leq 1/2$.
Since $\tilde{G}'(x\ra 0)=\ln 4-1>0 $  and $\tilde{G}'(1/2)=0$, we find  unique $x_c^{(1)}\in (0,1/2)$ such that $\tilde{G}'(x)>0$ for $x<x_c^{(1)}$, $\tilde{G}'(x)=0$ for $x=x_c^{(1)}$, and $\tilde{G}'(x)<0$ for $x<x_c^{(1)}$.
Consequently, $G(x)$ is monotonically increasing for  $0<x<x_c^{(1)}$ and decreasing for $x_c^{(1)}<x<1/2$.
Since $G(x\ra 0)=G(1/2)=0$, we have $G(x)\geq 0$ for $x\in (0,1/2]$.
}

{
Using \eqref{xlog} for every $\lambda_\alpha\:(1\leq \alpha\leq |R|)$ and take the sum, we obtain
\aln{
S_R\leq \sigma_{N_R}^2\ln y+\frac{4|R|}{ey}
}
for $y>0$.
}

\section*{Supplementary Note {4}: 
Proof of the bound in general quantum systems.}
Here, we assume that the observable is independent of time and show {(23)} in the main text.
We first note that the time derivative of the density matrix is generally given by 
$\fracd{\hat{\rho}}{t}=\frac{1}{2}\{\hat{\rho},\hat{L}\}$, where
$\hat{L}$ is the symmetric logarithmic derivative~\cite{PhysRevX.12.011038}.
Then, we have
\aln{
\fracd{\braket{\delta \hat{A}^2}}{t}
&=\frac{1}{2}\mr{Tr}[\delta\hat{A}^2\{\hat{\rho},\hat{L}\}]\nonumber\\
&=\frac{1}{2}\mr{Tr}[\delta\hat{A}\hat{\rho}\hat{L}\delta\hat{A}+\delta\hat{A}\hat{L}\hat{\rho}\delta\hat{A}]\nonumber\\
&=\frac{1}{2}\mr{Tr}\lrl{\delta\hat{A}\hat{\rho}\lrs{\hat{L}\delta\hat{A}-\fracd{\braket{{\hat{A}}}}{t}}+\lrs{\delta\hat{A}\hat{L}-\fracd{\braket{{\hat{A}}}}{t}}\hat{\rho}\delta\hat{A}}.
}
Using the Cauchy-Schwarz inequality,
we obtain
\aln{
\lrv{\fracd{\braket{\delta \hat{A}^2}}{t}}
&\leq 2\sqrt{\braket{\delta \hat{A}^2}}\sqrt{\frac{1}{4}\mr{Tr}[\delta\hat{A}\hat{L}\hat{\rho}\hat{L}\delta\hat{A}]-\frac{1}{4}\lrs{\fracd{\braket{\hat{A}}}{t}}^2}.
}
We thus have
\aln{
\lrs{\fracd{\braket{\hat{A}}}{t}}^2+4\lrs{\fracd{\sigma_A}{t}}^2\leq \braket{\hat{L}\delta\hat{A}^2\hat{L}}.
}
Using
\aln{
\braket{\hat{L}\delta\hat{A}^2\hat{L}}=\frac{\braket{\hat{L}\delta\hat{A}^2\hat{L}}}{\braket{\hat{L}^2}}\mathcal{F}_\mr{Q}\leq \|\delta\hat{A}\|_\infty^2\mathcal{F}_\mr{Q},
}
we obtain the bound in the main text.

{
We here also derive inequality (24) for unitary dynamics and general mixed states, assuming that $\hat{A}$ is time-independent.
We first note 
\aln{\label{energyvar}
&\mr{Tr}\lrs{i[\hat{H},\delta\hat{A}^2]\hat{\rho}}
=\mr{Tr}\lrs{i\delta\hat{A}\hat{\rho}\delta\hat{H}\delta\hat{A}-i\delta\hat{A}\delta\hat{H}\hat{\rho}\delta\hat{A}}\nonumber\\
&=\braketL{i\lrs{\delta\hat{H}\delta\hat{A}+\frac{i}{2}\fracd{\braket{{\hat{A}}}}{t}}\delta\hat{A}-i\delta\hat{A}\lrs{\delta\hat{A}\delta\hat{H}-\frac{i}{2}\fracd{\braket{{\hat{A}}}}{t}}}.
}
Using the Cauchy-Schwarz inequality, we have 
\aln{
\lrv{\fracd{\braket{\delta \hat{A}^2}}{t}}
&\leq 2\sigma_A{\sqrt{\mr{Tr}[\delta\hat{A}\delta\hat{H}\hat{\rho}\delta\hat{H}\delta\hat{A}]-\frac{1}{4}\lrs{\fracd{\braket{{\hat{A}}}}{t}}^2}}.
}
We thus have
\aln{
\lrs{\fracd{\braket{\hat{A}}}{t}}^2+4\lrs{\fracd{\sigma_A}{t}}^2\leq 4\braket{\delta \hat{H}\delta\hat{A}^2\delta\hat{H}}.
}
Finally, using the fact
\aln{
\braket{\delta \hat{H}\delta\hat{A}^2\delta\hat{H}}=\frac{\braket{\delta \hat{H}\delta\hat{A}^2\delta\hat{H}}}{\braket{\delta \hat{H}\delta\hat{H}}}\Delta E^2\leq \|\delta\hat{A}\|_\infty^2\Delta E^2,
}
we have the bound in the main text.
}

Let us specifically consider the case where $\hat{A}$ is a local observable  on a local subsystem X and $\hat{H}$ is a locally interacting Hamiltonian. In this case, $\delta\hat{A}^2$ also becomes a local observable on X.
Then, the left-hand side of Eq.~\eqref{energyvar} is replaced with
$\mr{Tr}\lrs{i[\hat{H}_\mathrm{X},\delta\hat{A}^2]\hat{\rho}}$, 
where $\hat{H}_\mathrm{X}$ is the Hamiltonian that consists of local interaction terms nontrivially acting on X.
Note that $\hat{H}_\mathrm{X}$ includes interactions at the boundary on X, which may include regions not contained in X.

Because of this replacement, we obtain
\aln{\label{unitary_local}
\lrs{\fracd{\braket{\hat{A}}}{t}}^2+4\lrs{\fracd{\sigma_A}{t}}^2&\leq 4\braket{\delta \hat{H}_\mr{X}\delta\hat{A}^2\delta\hat{H}_\mr{X}}\nonumber\\
&\leq 4\|\delta\hat{A}\|_\infty^2\Delta E_\mathrm{X}^2,
}
where $\Delta E_\mathrm{X}=\sigma_{H_\mr{X}}^2$. 
This expression means that for a local observable, its fluctuation and mean obey the tradeoff relation where the bound is given by the local energy fluctuation $\Delta E_\mr{X}$. 
While inequality~(24) in the main text typically becomes meaningless for many-body systems due to the divergence of $\Delta E$, inequality \eqref{unitary_local} provides a useful bound since $\Delta E_\mathrm{X}$ does not become large even if we increase the entire system size.

\section*{Supplementary Note {5}: 
Extension to multiple parameters.}
We here discuss how our bounds  based on the single parameter $t$ are generalized to the multi-parameter case.
This is relevant when we want to evaluate the rate of the change of the fluctuation of an observable in response to the change of those parameters.

For this purpose, we consider a state depending on $M$ parameters, $\vec{\lambda}=(\lambda_1,\cdots,\lambda_Z)$.
For simplicity, let us consider an observable ${A}$ independent of $\vec{\lambda}$ and define the vectors 
\aln{
\vec{\mathsf{S}}_A=\lrs{\fracpd{\sigma_{A}}{\lambda_1},\cdots,\fracpd{\sigma_{A}}{\lambda_Z}}^\mathsf{T}
}
and
\aln{
\vec{\mathsf{B}}_A=\lrs{\fracpd{\braket{A}}{\lambda_1},\cdots,\fracpd{\braket{A}}{\lambda_Z}}^\mathsf{T}.
}
We then find the following matrix inequalities as a generalization of the single-parameter case:
{
\aln{\label{matineq}
\vec{\mathsf{S}}_A\vec{\mathsf{S}}_A^\mathsf{T}\preceq c\:\mathsf{cov}(\vec{\mc{V}}_A,\vec{\mc{V}}_A)
}
and
\aln{\label{matineq2}
\vec{\mathsf{B}}_A\vec{\mathsf{B}}_A^\mathsf{T}+\frac{1}{c^2}\vec{\mathsf{S}}_A\vec{\mathsf{S}}_A^\mathsf{T}\preceq \mathsf{C}_A, 
}
}
where
\aln{
\vec{\mc{V}}_A=(\mc{V}_{A,1},\cdots,\mc{V}_{A,Z})^\mathsf{T}
}
denotes a set of velocity observables
satisfying 
\aln{
\fracpd{\braket{A}}{\lambda_z}=\braket{\mc{V}_{A,z}},
}
\aln{
(\mathsf{C}_A)_{z,z'}=\braket{\mc{V}_{A,z},\mc{V}_{A,z'}}
}
is the matrix relevant for the tradeoff,
and
\aln{
[\mathsf{cov}(\vec{\mc{V}}_A,\vec{\mc{V}}_A)]_{z,z'}
=\braket{\mc{V}_{A,z},\mc{V}_{A,z'}}-\braket{\mc{V}_{A,z}}\braket{\mc{V}_{A,z'}}
=(\mathsf{C}_A)_{z,z'}-(\vec{\mathsf{B}}_A)_z(\vec{\mathsf{B}}_A)_{z'}
}
is the covariance matrix.
The matrix inequality $X\preceq Y$ means that $Y-X$ is positive semidefinite.
We note that \eqref{matineq} indicates a scalar inequality, $\vec{\mathsf{S}}_A^\mathsf{T}[\mathsf{cov}(\vec{\mc{V}}_A,\vec{\mc{V}}_A)]^{-1}\vec{\mathsf{S}}_A\leq 1$, provided that $\mathsf{cov}(\vec{\mc{V}}_A,\vec{\mc{V}}_A)$ has an inverse.
We also note that the above inequalities reduce to those seen in the main text if we consider $Z=1$.

Let us first show the above bounds using the conservation law of probability for each parameter's change in continuous systems.
Given $\rho(\vec{\lambda})$, we assume
\aln{
\fracpd{\rho}{\lambda_z}=-\nabla\cdot({\rho\mbf{V}_z})
}
for every $z=1,\cdots,Z$.
Let's define $\mc{V}_{A,z}=\nabla A\cdot\mbf{V}_z$.
Then, for any rea
 l nonzero vector $\vec{a}_z=(a_1,\cdots,a_Z)^\mathsf{T}$,
we have (see Ref.~\cite{hamazaki2023quantum} for a similar technique used for speed limits in the mean)
\aln{
\sum_za_z\fracpd{\braket{\delta A^2}}{\lambda_z}=2\braketL{\delta A\sum_za_z\delta\mc{V}_{A,z}}
}
Taking the square, we have
\aln{
\sum_{zz'}a_za_{z'}4\sigma_A^2\fracpd{\sigma_A}{\lambda_z}\fracpd{\sigma_A}{\lambda_{z'}}\leq 4\sigma_A^2
\sum_{zz'}a_za_{z'}\braketL{\delta\mc{V}_{A,z}\delta\mc{V}_{A,z'}}.
}
Since this holds for every $\vec{a}$, we find the inequality in \eqref{matineq}, and thus \eqref{matineq2}, {with $c=1$}.
We also note that, since
\aln{
\sum_{zz'}a_za_{z'}\braketL{\mc{V}_{A,z}\mc{V}_{A,z'}}
\leq\|\nabla A\|_\infty^2\sum_{zz'}a_za_{z'}
\braketL{\mbf{V}_{z}\cdot\mbf{V}_{z'}}
}
and thus
\aln{
\mathsf{C}_A\preceq \|\nabla A\|_\infty^2 \mathsf{W}
}
with $\mathsf{W}_{zz'}=\braketL{\mbf{V}_{z}\cdot\mbf{V}_{z'}}$.
We have considered continuous systems above, but a similar discussion can be done for discrete systems described in \textbf{Methods: Extension of the current-based bound to discrete systems}.

{Second}, we consider the unitary quantum evolution, where we assume
\aln{
\fracpd{\hat{\rho}}{\lambda_z}=-i[\hat{H}_z,\hat{\rho}]
}
for Hamiltonians $\hat{H}_1,\cdots,\hat{H}_Z$.
{Again, by considering 
\aln{
\sum_za_z\fracpd{\braket{\delta \hat{A}^2}}{\lambda_z}=2\braketL{\delta \hat{A},\sum_za_z\delta\hat{\mc{V}}_{A,z}},
}
}
with $\hat{\mc{V}}_{A,z}=i[\hat{H}_z,\hat{A}]$, we find \eqref{matineq} and \eqref{matineq2} {for  
$c=1$}.

{Third}, we consider the information-theoretical bound for discrete classical systems.
By setting $(\mc{V}_{A,z})_i=\delta A_i\frac{1}{p_i} \fracpd{p_i}{\lambda_z}$, we find 
\aln{
\sum_za_z\fracpd{\braket{\delta A^2}}{\lambda_z}=\braketL{\delta A\sum_za_z\delta\mc{V}_{A,z}}
}
 for any real nonzero vector $\vec{a}_z=(a_1,\cdots,a_Z)^\mathsf{T}$.
We then have
\aln{
\vec{\mathsf{S}}_A\vec{\mathsf{S}}_A^\mathsf{T}\preceq \frac{1}{4}\mathsf{cov}(\vec{\mc{V}}_A,\vec{\mc{V}}_A)
}
and 
\aln{\label{fisq}
\vec{\mathsf{B}}_A\vec{\mathsf{B}}_A^\mathsf{T}+4\vec{\mathsf{S}}_A\vec{\mathsf{S}}_A^\mathsf{T}\preceq \mathsf{C}_A, 
}
{which correspond to \eqref{matineq} and \eqref{matineq2} with $c=1/2$.}
Furthermore, we find
\aln{\label{fisq2}
\mathsf{C}_A\preceq \|\delta A\|_\infty^2 \mathsf{F}_C, 
}
where
$(\mathsf{F}_C)_{zz'}=\sum_i\frac{1}{p_i}\fracpd{p_i}{\lambda_z}\fracpd{p_i}{\lambda_{z'}}$ is the classical Fisher information matrix~\cite{cover1999elements}.
Thus, as in the single-parameter case, {the upper bound is described} using the information content.

Finally, extending Supplementary Note 4, we consider general quantum systems and obtain the information-theoretical bound.
We assume 
\aln{
\fracpd{\hat{\rho}}{\lambda_z}=\frac{1}{2}\{\hat{L}_z,\hat{\rho}\},
}
where $\hat{L}_z$ is the symmetric logarithmic derivative for the parameter $\lambda_z$.
We then have
\aln{
\lrv{\sum_za_z\fracpd{\braket{\delta \hat{A}^2}}{\lambda_z}}
&=\lrv{\sum_za_z\frac{1}{2}\mr{Tr}[\delta\hat{A}^2\{\hat{\rho},\hat{L}_z\}]}\nonumber\\
&=\frac{1}{2}\lrv{\mr{Tr}\lrl{\delta\hat{A}\hat{\rho}\sum_za_z\lrs{\hat{L}_z\delta\hat{A}-\fracpd{\braket{{\hat{A}}}}{\lambda_z}}+\sum_za_z\lrs{\delta\hat{A}\hat{L}_z-\fracpd{\braket{{\hat{A}}}}{\lambda_z}}\hat{\rho}\delta\hat{A}}}\nonumber\\
&\leq\frac{1}{2}\lrv{\mr{Tr}\lrl{\delta\hat{A}\hat{\rho}\sum_za_z\lrs{\hat{L}_z\delta\hat{A}-\fracpd{\braket{{\hat{A}}}}{\lambda_z}}}}+\frac{1}{2}\lrv{\mr{Tr}\lrl{{\sum_za_z\lrs{\delta\hat{A}\hat{L}_z-\fracpd{\braket{{\hat{A}}}}{\lambda_z}}\hat{\rho}\delta\hat{A}}}}\nonumber\\
&\leq\sqrt{\braket{\delta \hat{A}^2}\sum_{zz'}a_za_{z'}\braketL{\lrs{\hat{L}_z\delta\hat{A}-\fracpd{\braket{{\hat{A}}}}{\lambda_z}}\lrs{\delta\hat{A}\hat{L}_{z'}-\fracpd{\braket{{\hat{A}}}}{\lambda_{z'}}}}}
}
Taking the square and using the Cauchy-Schwarz inequality,
we have
\aln{
\sum_{zz'}a_za_{z'}\fracpd{\sigma_A}{\lambda_z}\fracpd{\sigma_A}{\lambda_{z'}}
\leq& \frac{1}{4}\sum_{zz'}a_za_{z'}
\braketL{\lrs{\hat{L}_z\delta\hat{A}-\fracpd{\braket{{\hat{A}}}}{\lambda_z}}\lrs{\delta\hat{A}\hat{L}_{z'}-\fracpd{\braket{{\hat{A}}}}{\lambda_{z'}}}}\nonumber\\
=& \frac{1}{4}\sum_{zz'}a_za_{z'}
\lrs{\braket{\hat{L}_z\delta\hat{A}^2\hat{L}_{z'}}-\fracpd{\braket{\hat{A}}}{\lambda_z}\fracpd{\braket{\hat{A}}}{\lambda_{z'}}}
\nonumber\\
\leq&\frac{1}{4}\sum_{zz'}a_za_{z'}
\lrs{\|\delta\hat{A}\|_\infty^2(\mathsf{F}_Q)_{zz'}-\fracpd{\braket{\hat{A}}}{\lambda_z}\fracpd{\braket{\hat{A}}}{\lambda_{z'}}},
\nonumber\\
}
where $(\mathsf{F}_Q)_{zz'}=\braket{\hat{L}_z,\hat{L}_{z'}}$ is the quantum Fisher information matrix.

Thus, we obtain
\aln{
\vec{\mathsf{B}}_A\vec{\mathsf{B}}_A^\mathsf{T}+4\vec{\mathsf{S}}_A\vec{\mathsf{S}}_A^\mathsf{T}\preceq \mathsf{C}_A'\preceq\|\delta\hat{A}\|_\infty^2\:\mathsf{F}_Q
}
with $(\mathsf{C}_A')_{zz'}=\braket{\hat{L}_z\delta\hat{A}^2\hat{L}_{z'}}$.
This is a quantum version of \eqref{fisq} and \eqref{fisq2}.

{We note that we can also obtain the corresponding bound for the unitary dynamics,}
\aln{
\vec{\mathsf{B}}_A\vec{\mathsf{B}}_A^\mathsf{T}+4\vec{\mathsf{S}}_A\vec{\mathsf{S}}_A^\mathsf{T}\preceq \mathsf{C}_A'\preceq\|\delta\hat{A}\|_\infty^2\:\mathsf{cov}\lrs{\vec{\hat{H}},\vec{\hat{H}}}, 
}
where
\aln{
(\mathsf{C}_A')_{zz'}=\braket{\delta\hat{H}_z\delta\hat{A}^2\delta\hat{H}_{z'}}
}
and
\aln{
\lrl{\mathsf{cov}\lrs{\vec{\hat{H}},\vec{\hat{H}}}}_{zz'}=\braket{\hat{H}_z,\hat{H}_{z'}}-\braket{\hat{H}_z}\braket{\hat{H}_{z'}}.
}
Thus, {the upper bound is described with} the Hamiltonian covariance.


{
\section*{Supplementary Note 6: 
The alternative derivation of Eq. (10) in the main text.}
We here show that Eq. (10) can also be derived from the known results in Wasserstein distance~\cite{dechant2019thermodynamic,PhysRevResearch.3.043093}, i.e.,
\aln{\label{entsup}
{\lim_{\delta t\ra 0}\frac{\mc{W}_2(\rho(t),\rho(t+\delta t))^2}{\delta t^2}}\leq \mu T\dot{\Sigma}.
}
}

{
For this purpose, we remind of the definition of the order-2 Wasserstein distance,
\aln{\label{Pim}
\mc{W}_2(\rho,\nu)^2=\min_\Pi \int d\mbf{x}\int d\mbf{y} |\mbf{x}-\mbf{y}|^2 \Pi(\mbf{x},\mbf{y}),
}
where $\Pi(\mbf{x},\mbf{y})$ is a coupling between $\rho$ and $\nu$, i.e., $\int d\mbf{x} \Pi(\mbf{x},\mbf{y})=\nu(\mbf{y})$ and 
$\int d\mbf{y} \Pi(\mbf{x},\mbf{y})=\rho(\mbf{x})$.
We will consider $\rho=\rho(t)$ and $\nu=\rho(t+\delta t)$ in the end.
}

{
Let  $\Pi_m(\mbf{x},\mbf{y})$ be the optimal coupling the achieves the minimization in Eq.~\eqref{Pim}.
We then have
\aln{
\int d\mbf{x}\int d\mbf{y} |A(\mbf{x})-A(\mbf{y})|^2 \Pi_m(\mbf{x},\mbf{y})
&\leq \|\nabla A\|_\infty^2\int d\mbf{x}\int d\mbf{y} |\mbf{x}-\mbf{y}|^2 \Pi_m(\mbf{x},\mbf{y})\nonumber\\
&=\|\nabla A\|_\infty^2\mc{W}_2(\rho,\nu)
}
because $\|\nabla A\|_\infty^2$ serves as a Lipschitz constant for $A$.
}

{
Furthermore, the left-hand side is given by
\aln{
\int d\mbf{x}\int d\mbf{y} |A(\mbf{x})-A(\mbf{y})|^2 \Pi_m(\mbf{x},\mbf{y})
&= \braket{A^2}_\rho +\braket{A^2}_\nu- 2\int d\mbf{x}\int d\mbf{y} A(\mbf{x})A(\mbf{y}) \Pi_m(\mbf{x},\mbf{y})\nonumber \\
&=(\braket{A}_\rho-\braket{A}_\nu)^2
+\sigma_{A,\rho}^2 +\sigma_{A,\mu}^2\nonumber\\
&\:\:\:- 2\int d\mbf{x}\int d\mbf{y} \delta A(\mbf{x})\delta A(\mbf{y}) \Pi_m(\mbf{x},\mbf{y})
}
where we write the expectation value and the standard deviation for $A$ with respect to a probability distribution $\rho$ as $\braket{A}_\rho$ and $\sigma_{A,\rho}$, respectively.
Finally, we use the Cauchy-Schwarz inequality to find
\aln{
\int d\mbf{x}\int d\mbf{y} |A(\mbf{x})-A(\mbf{y})|^2 \Pi_m(\mbf{x},\mbf{y})
&\geq (\braket{A}_\rho-\braket{A}_\nu)^2
+\sigma_{A,\rho}^2 +\sigma_{A,\mu}^2-2\sqrt{\braket{\delta A^2}_\rho \braket{\delta A^2}_\nu}\nonumber\\
&=(\braket{A}_\rho-\braket{A}_\nu)^2+(\sigma_{A,\rho}-\sigma_{A,\nu})^2
}
}

{
Therefore, we have
\aln{
(\braket{A}_\rho-\braket{A}_\nu)^2+(\sigma_{A,\rho}-\sigma_{A,\nu})^2
\leq \|\nabla A\|_\infty^2\mc{W}_2(\rho,\nu).
}
We finally use this assuming $\rho=\rho(t)$ and $\nu=\rho(t+\delta t)$, take $\delta t\ra 0$ limit, and remember \eqref{entsup} to find the general-observable version of Eq. (10),
\aln{
\lrs{\fracd{\braket{A}}{t}}^2+\lrs{\fracd{\sigma_{A}}{t}}^2\leq \|\nabla A\|_\infty^2 \mu T\dot{\Sigma}
}
}

\bibliographystyle{naturemag}
\bibliography{reference}